\documentclass[10pt,a4paper,twocolumn,english,prb,aps,showpacs,floatfix,groupedaddress]{revtex4-1}

\usepackage{graphicx}
\usepackage{epsfig}
\usepackage[english]{babel}
\usepackage{amsmath}
\usepackage{amssymb}
\usepackage{amsfonts}
\usepackage{longtable}
\usepackage{dcolumn}
\usepackage{bm}
\usepackage{bbm}
\usepackage{nicefrac}
\usepackage{color,array}
\usepackage{colortbl}

\usepackage{dsfont}

\usepackage{ulem}

\def\jq{J_\text{Q}}
\def\R{_\text{R}}
\def\neff{N_\text{eff}}
\def\ntr{N_\text{tr}}

\begin{document}
\title{Nuclear frequency focusing in periodically pulsed
semiconductor quantum dots described
by infinite classical central spin models}

\author{Philipp Schering}
\email{philipp.schering@tu-dortmund.de}
\affiliation{Lehrstuhl f\"{u}r Theoretische Physik I, Technische Universit\"at Dortmund, Otto-Hahn Stra\ss{}e 4, 44221 Dortmund, Germany}

\author{Benedikt Fauseweh}
\email{b.fauseweh@fkf.mpg.de}
\affiliation{Max-Planck Institut f\"ur Festk\"orperforschung, Heisenbergstra\ss{}e 1, 
70569 Stuttgart, Germany}

\author{Jan H\"udepohl}
\affiliation{Lehrstuhl f\"{u}r Theoretische Physik I, Technische Universit\"at Dortmund, Otto-Hahn Stra\ss{}e 4, 44221 Dortmund, Germany}

\author{G\"otz S.\ Uhrig}
\email{goetz.uhrig@tu-dortmund.de}
\affiliation{Lehrstuhl f\"{u}r Theoretische Physik I, Technische Universit\"at Dortmund, Otto-Hahn Stra\ss{}e 4, 44221 Dortmund, Germany}

\date{\rm\today}

\begin{abstract}
The coherence of an electronic spin in a semiconductor quantum dot decays due
to its interaction with the bath of nuclear spins in the surrounding
isotopes. This effect can be reduced by subjecting the system to 
an external magnetic field and by applying optical pulses.
By repeated pulses in long trains the spin precession can be 
synchronized to the pulse period $T_\text{R}$.
This drives the nuclear spin bath into states far from equilibrium leading to nuclear frequency
focusing. In this paper, we use an efficient classical approach introduced in 
Phys.\ Rev.\ B {\bf 96}, 054415 (2017) to describe and to analyze this nuclear 
focusing. Its dependence on the effective bath size and on the external magnetic field 
 is elucidated in a comprehensive study. We find that the characteristics of the pulse
as well as the nuclear Zeeman effect influence the behavior decisively.
\end{abstract}

\pacs{03.65.Yz, 78.67.Hc, 72.25.Rb, 03.65.Sq}

\maketitle

\section{Introduction}
\label{sec:intro}

For almost 20 years the electronic spin in quantum dots is considered a promising
candidate for the realization of quantum bits \cite{loss98,schli03} which
are at the very basis of any quantum information processing \cite{niels00}. Considerable effort
has been invested in the experimental investigation of the spin dynamics in semiconductor
nano-structures and the possibilities to manipulate it \cite{hanso07,urbas13}.
It is particularly interesting  that ensembles of quantum dots
can be manipulated as well. They can be made to respond coherently by subjecting them 
to long periodic trains of optical pulses \cite{greil06a,greil06b,greil07a}. 
It appears that the periodic pulsing with repetition time $T_\text{R}$ synchronizes
the Larmor precessions of the spins in sub-ensembles of quantum dots. The 
Overhauser field, i.\,e., the magnetic field applied by the nuclear spins via
hyperfine coupling on the electronic spin, 
changes such that it compensates the fluctuations in the $g$ factor
from dot to dot which otherwise would lead to fast dephasing of the Larmor precessions
of different dots. This phenomenon is called nuclear frequency focusing.

It is a particular challenge to understand the dynamics of nuclear frequency focusing. 
The fundamental model is the central spin model (CSM) where the central spin 
stands for the electronic (or hole) spin while the bath spins represent the nuclear spins.
This model was first considered and analytically solved in the
stationary case, i.\,e., without any pulsing, by Gaudin \cite{gaudi83}.
The evaluation, however, of the ensuing Bethe ansatz equations is far from trivial 
\cite{farib13a}.

There is a growing number of theoretical studies which are devoted to the issue
of periodic pulsing of the CSM \cite{barne11b,petro12,yugov12,glazov12,econo14,beuge16}. 
The task is difficult because the number $\neff$ of relevant nuclear spins 
in the bath is as large as $10^5$, which renders the relevant Hilbert spaces
{intractably} large\cite{merku02,schli03,lee05,petro08}. 
The relevant time scales to be described are very large as well
and reach seconds, if not minutes, while the intrinsic time scale of the dynamics
of the electronic spin is in the 
range of nanoseconds. The bath size and the time scales are way beyond what
can be tackled in theoretical calculations by standard tools
such as exact diagonalization \cite{cywin10}, Chebyshev expansion \cite{dobro03a}, or 
density-matrix renormalization \cite{stane13}.

It was shown that simulations of the classical CSM
averaged suitably over initial conditions reproduce the quantum mechanical
solutions very well \cite{stane14b}. This behavior can be justified by the
large number of contributing bath spins \cite{stane14b} and by path-integral
arguments \cite{alhas06,chen07}. But it has been ascertained recently
that the limit $\neff\to\infty$ does not lead to 
a purely classical model, although the numerical results from 
the quantum mechanical and the classical calculations are
very close to each other \cite{rohri18}.

However, the computational task is extremely challenging even on the classical level. 
A very efficient algorithm for the
simulation of infinitely large classical spin baths has been introduced 
\cite{fause17a} which renders the required bath sizes tractable.
Also large times can be addressed as well even though one cannot reach
the experimental scales. A very useful observation in this context is
that the intrinsic time scale of the spin bath scales like the square root of the effective number $\neff$ of coupled bath spins, i.\,e.,
$t_\text{scale} \propto \sqrt{\neff} /\jq$, where
we set $\hbar=1$ and $J_\text{Q}$ is the energy scale of the central spin dynamics.
Thus, one can perform calculations for smaller baths and scale them up to
the orders of magnitude relevant in experiment. Whether such scaling still
holds in pulsed systems in magnetic fields is one of the open issues
in the field.

The objective of the present paper is to make use of the {methodological}
progress to consider the effect of periodic pulsing using the improved
approaches. In order to keep the simulations simple and efficient,
we use approximate classical pulses similar to the pulse considered in 
Ref.\ \onlinecite{petro12}.
This pulse aligns the central spin into the $z$-direction independent of the 
direction it had before the pulse. Thus, the alignment induced by the 
pulse is perpendicular to the applied external magnetic field which
is oriented along the $x$-direction (Voigt geometry).

The paper is set up as follows. In Sect.\ \ref{sec:modsim}
we introduce the model and the equations of motion to be solved,
we discuss the pulses considered, and we recapitulate the 
efficient algorithm used for the simulations. 
In Sect.\ \ref{sec:NNZ} we provide representative results for the 
CSM subject to periodic pulsing without coupling
of the bath spins to the external magnetic field, i.\,e., neglecting
the nuclear Zeeman term. Clear evidence for nuclear frequency focusing is found.
The scaling behavior with respect to the size of the spin bath and to the applied 
magnetic field is studied. In Sect.\ \ref{sec:WNZ}, we
consider the effects of the additional Larmor precession of the nuclear
spins about the external magnetic field, which have turned out to
be relevant very recently \cite{beuge17}. 
The conclusions are presented in Sect.\ \ref{sec:conclusion}.

\section{Model and simulation}
\label{sec:modsim}

\subsection{Model}
\label{ssec:model}

The Hamilton function of the classical CSM to be considered reads
\begin{equation}
\label{eq:hamil}
H= \vec{S}_0 \sum_{k=1}^N J_k\vec{S}_k - h S_0^x- z h \sum_{k=1}^N S^x_k,
\end{equation}
where the vector $\vec{S}_0$ stands for the central, electronic spin
and the vectors $\vec{S}_k$ stand for the nuclear spins forming the bath.
The hyperfine couplings $J_k$ represent the coupling between electronic
and bath spins. The external magnetic field $h=g\mu_\text{B} B$
is applied in Voigt geometry along the $x$-direction. A generic electronic 
$g$-factor is $0.555$ \cite{greil07a}.
The last term in Eq.\ \eqref{eq:hamil} is the nuclear Zeeman term where
$z=g_\text{nuclear}/g$ implements the reduced nuclear $g$-factor; a generic 
value is $z=1/800$ \cite{beuge17}.

The couplings $J_k$ are proportional to the probability of finding the electron
at the site of the nucleus. For simplicity, we assume that the couplings
can be parameterized exponentially \cite{merku02,schli03,farib13b,fause17a}
\begin{equation}
\label{eq:couplings}
J_k =C\exp(-k\gamma),
\end{equation}
where $C$ is an energy constant and $\gamma\ll 1$ is a small parameter given by
$2/\neff$ with $\neff$ being the number of effectively coupled
bath spins. Note that the total number of bath spins is infinite $N\to\infty$
because all spins in the sample are coupled to
the central spin, though perhaps extremely weakly.
The number $\neff$ quantifies the number of bath spins which 
are appreciably coupled to the central spin, for details see
Ref.\ \onlinecite{fause17a}. Often they are referred to as the nuclear spins within
the localization volume of the electronic wave function.
Hence, a realistic value \cite{coish09} for $\gamma$ is $10^{-4}$ to $10^{-6}$.

The constant $C$ in \eqref{eq:couplings} is specified via the square root of the sum of the
squared couplings
\begin{equation}
\label{eq:jq}
J_\text{Q} =\sqrt{\sum_{k=1}^N J_k^2}
\end{equation}%
because it is this energy which determines the rate of the central spin
dynamics for short times. For small values of $\gamma$,
one finds $C=\sqrt{2\gamma}J_\text{Q}$ \cite{fause17a}. 
The generic range of $J_\text{Q}$ is between
$1$ and $10$ $\mu$eV corresponding to a time constant of about $1\,\mathrm{ns}$.

The Overhauser field $\vec{B}$ is given by the weighted sum of all
bath spins
\begin{equation}
\label{eq:overhaus}
\vec{B} =\sum_{k=1}^N J_k \vec{S}_k.
\end{equation}
Since we assume that the initial bath is completely disordered,
we describe it by randomly chosen initial configurations.
The variance of any of the components $B^\alpha$ ($\alpha\in\{x,y,z\}$) of
the Overhauser field is set to its quantum mechanical values, i.\,e.,
\begin{equation}
\label{eq:overhaus_variance}
\overline{(B^\alpha)^2} = \frac{1}{Z}\text{Tr} ((\hat B^\alpha)^2) = \frac{5J^2_\text{Q}}{4},
\end{equation}
where the factor $5/4$ follows from the observation that the bath spins 
\cite{walch55,coish09,stone15,beuge17} have $S=3/2$
if no indium needs to be considered. Clearly, this could be 
changed easily to other values if considered appropriate.

\subsection{Simulation}
\label{ssec:simul}

The equations of motion resulting from \eqref{eq:hamil} are the 
well-known differential equations describing precessions
\begin{align}
\label{eq:exact_central_spin}
\frac{\mathrm{d}}{\mathrm{d}t} \vec{S}_0 = \left(\vec{B}-\vec{h}\right) \times \vec{S}_0 ,
\end{align}
where $\vec{h}:=(h,0,0)^\text{T}$, for the central spin and
\begin{align}
\label{eq:exact_bath_spin}
\frac{\mathrm{d}}{\mathrm{d}t} \vec{S}_k = \left(J_k \vec{S}_0 - z\vec{h}\right) 
\times \vec{S}_k
\end{align}
for each bath spin. The factor $z\approx 10^{-3}$ takes into account
that the nuclear magnetic moment is three orders smaller than
the electronic one. While these equations can be numerically
solved by standard algorithms such as the Runge-Kutta algorithm
of various orders, the direct simulation of $10^5$ equations, let alone
of an infinite number of them, is not an option. 

Thus, we resort to the spectral density approach introduced 
previously \cite{fause17a}. The ensemble of bath spins parametrized 
according to \eqref{eq:couplings} can be represented by the linear weight function 
$W(\varepsilon)=(\varepsilon/\gamma)\theta(\sqrt{2\gamma}-\varepsilon)$
where we set the energy scale $\jq$ to unity. The energy range
$[0,\sqrt{2\gamma}]$ is divided into $\ntr$ intervals 
$I_i:=[\tilde\epsilon_{i+1},\tilde\epsilon_i]$. The most efficient choices
are intervals which become exponentially small for increasing $i$ (see Ref.\ \onlinecite{fause17a}). 
We choose
\begin{equation}
\label{eq:interval-def}
\tilde\epsilon_i = \lambda^i \sqrt{2\gamma}\left(\frac{\ntr-i}{\ntr}\right),
\quad i\in\{0,1,2,\ldots,\ntr\}
\end{equation}
and determine $\lambda$ from
\begin{equation}
\lambda = \left(\frac{\ntr}{\sqrt{2\gamma}t_\text{max}} \right)^{1/(\ntr-1)},
\end{equation}
where $t_\text{max}$ is the maximum time up to which the simulation will
be performed \cite{fause17a}.

Within each interval, $W_i$ defines the weight
\begin{equation}
W_i := \int_{\tilde\epsilon_{i}}^{\tilde\epsilon_{i-1}} W(\varepsilon)d\varepsilon
\end{equation}
and $\varepsilon_i$ defines the average energy 
\begin{equation}
\varepsilon_i := \frac{1}{W_i}\int_{\tilde\epsilon_{i}}^{\tilde\epsilon_{i-1}} \varepsilon
W(\varepsilon)d\varepsilon,
\quad i\in\{1,2,\ldots,\ntr\}.
\end{equation}
The sum of the spin operators of the bath in which the couplings
lie within the interval $I_i$ defines the vector $\vec{Q}_i$.
This vector obeys the equation of motion
\begin{equation}
\frac{\mathrm{d}}{\mathrm{d}t} \vec{Q}_i = \left( \varepsilon_i \vec{S}_0-z\vec{h}\right)
\times \vec{Q}_i. 
\end{equation}
{It is chosen randomly with fixed variance, see below for details. The 
differing contribution of the $\vec{Q}_i$
to the Overhauser field $\vec{B}$ is accounted for by the square roots
of the weights of the $W_i$}
\begin{equation}
\vec{B} = \sum_{i=1}^{\ntr} \sqrt{W_i} \vec{Q}_i.
\end{equation}
{For the derivation we refer the reader to Ref.\ \onlinecite{fause17a}.}
The key advantage of this approach is that the number $3\ntr$ of equations 
to be followed ranges only in the hundreds instead of $10^5$. The convergence
is uniform with $\ntr$. In the time interval for which the
discretization is optimized the accuracy is essentially the same for all times.
The deviations decrease quadratically upon increasing $\ntr$. 
It is this approach which we employ in the present paper \cite{fause17a}.

Since we intend to use the classical simulation as approximation
of the quantum mechanical problem, we average over the initial configurations
to determine the average autocorrelation function 
\begin{align}
\label{eq:measure_def}
S^{zz}(t) &:= \overline{S_0^z(t)S_0^z(0)}
\end{align}
as approximation to the quantum mechanical autocorrelation function 
$\langle S_0^z(t)S_0^z(0)\rangle$. This procedure has proven to yield
very reasonable results \cite{stane14b,rohri18} and hence we use it here as well.
Each spin component of each bath spin and of the central spin is chosen
according to Gaussian distribution functions centered around zero and with variance
$1/4$ for the central spin, being $S=1/2$, and with variance $5/4$ for the bath spins, being $S=3/2$.

Since the fields $\vec{Q}_i$ are linear sums of the bath spins,
their components are also Gaussian distributed.
They are uncorrelated for different intervals $j$ and $m$.
 Hence, at $t=0$, we have
\begin{align}
\overline{Q_j^\alpha Q_m^\beta} &= \delta_{j,m}\delta_{\alpha,\beta} \frac{5}{4}
\end{align}
so that we can directly initialize the $3N_\text{tr}$ fields $Q_j^\alpha$
according to Gaussian distributions.
{We stress that the same variance for all $Q_j^\alpha$ can be chosen
because we consider a single species of nuclear spins, here $I=3/2$.
The influence of the differing numbers of bath spins \cite{fause17a} contributing 
to each $\vec{Q}_j$ is accounted for by the weights $W_j$.}

Next, we turn to the pulses acting on the central spin.
In real experiments, the optical pulses of circularly polarized light 
of well-defined frequency excite trions of only one spin orientation which decay quickly,
leaving behind a (partially) polarized spin \cite{greil06a,yugov09,yugov12,glazov12,jasch17}. The trion decays fast, though not instantaneously, on the time
scale of $0.4\,\mathrm{ns}$. We neglect this time and mimic the whole pulse
by an orientation of the central spin along the $z$-axis.

As a first {description of the} pulse, we consider pulse model I
\begin{align}
\label{eq:pulse1}
\vec{S}_0 &\to \begin{pmatrix}
0 \\ 0 \\ |\vec{S}_0|
\end{pmatrix}
\end{align}
which rotates the full vector $\vec{S}_0$ into the $z$-axis while preserving its length.
This pulse is idealized in the sense that it does not respect Heisenberg's 
uncertainty relation for the central spin. Pulse model I is very close to 
the classical pulse studied in Ref.\ \onlinecite{petro12} by Petrov and Yakovlev.

As a second pulse, we consider pulse model II, which is set up
to mimic the quantum mechanical aspects better. The guiding
idea is that the pulse represents a quantum mechanical measurement
with the outcome of a spin up. In order to realize this 
behavior, we choose the spin after the pulse to be
\begin{align}
\label{eq:pulse2}
\vec{S}_0 &\to \begin{pmatrix}
X \\
Y \\
1/2 \end{pmatrix},
\end{align}
where $X$ and $Y$ are chosen for each pulse at random from a Gaussian ensemble
with variance $1/4$. This implies vanishing expectation values of $X$ and $Y$,
and the spin length is correct on average.

Alternatively, $X$ and $Y$ could be chosen uniformly distributed on a circle with radius 
$X^2+Y^2=(1/2)^2$, which could be a third pulse. However, we tested that 
the difference to pulse model II is hardly noticeable, and hence we restrict 
ourselves to pulse model II in addition to pulse model I.
We expect that pulse model I is more efficient in generating nuclear focusing, but that 
pulse model II is more realistic in mimicking the quantum mechanical system better.

\section{Results for the pulsed system without nuclear Zeeman coupling}
\label{sec:NNZ}

In this section, we neglect the nuclear Zeeman term in the Hamiltonian 
\eqref{eq:hamil}, i.\,e., we set $z=0$.
This allows us to study the influence of this term later in {Sect.} 
\ref{sec:WNZ} by comparing the different results with and without the nuclear Zeeman term.

Typical experiments are done at external magnetic fields of 
$1$ to $6\,\mathrm{T}$.
In our units, $1\,\mathrm{T}$ corresponds approximately {to} $h = 40\jq$.

All calculations are averaged over $10^5$ initial configurations,
except those for  representative illustrations. The data for 
the representative illustrations {are} averaged over $10^6$ configurations.

The truncation parameter is chosen as $N_\mathrm{tr} = 44$ for simulations up to 
$n_\mathrm{p} = 10^4$ pulses. 
For longer simulations, we increase $N_\mathrm{tr}$ such that the discretization parameter 
$\lambda$ used in \eqref{eq:interval-def} remains constant. This ensures the same
accuracy level independent of the number of pulses studied.
Note that due to the exponential discretization of the weight function, the 
truncation parameter has to be increased only slighty even for much longer simulations, 
e.\,g., $N_\mathrm{tr} = 58$ is sufficient for ten times more pulses, $n_\mathrm{p} = 10^5$.

\subsection{Pulse model I from Eq. \eqref{eq:pulse1}}
\label{subsec:NNZ_1}

\begin{figure}[htb]
\centering
\includegraphics[width=1.0\columnwidth]{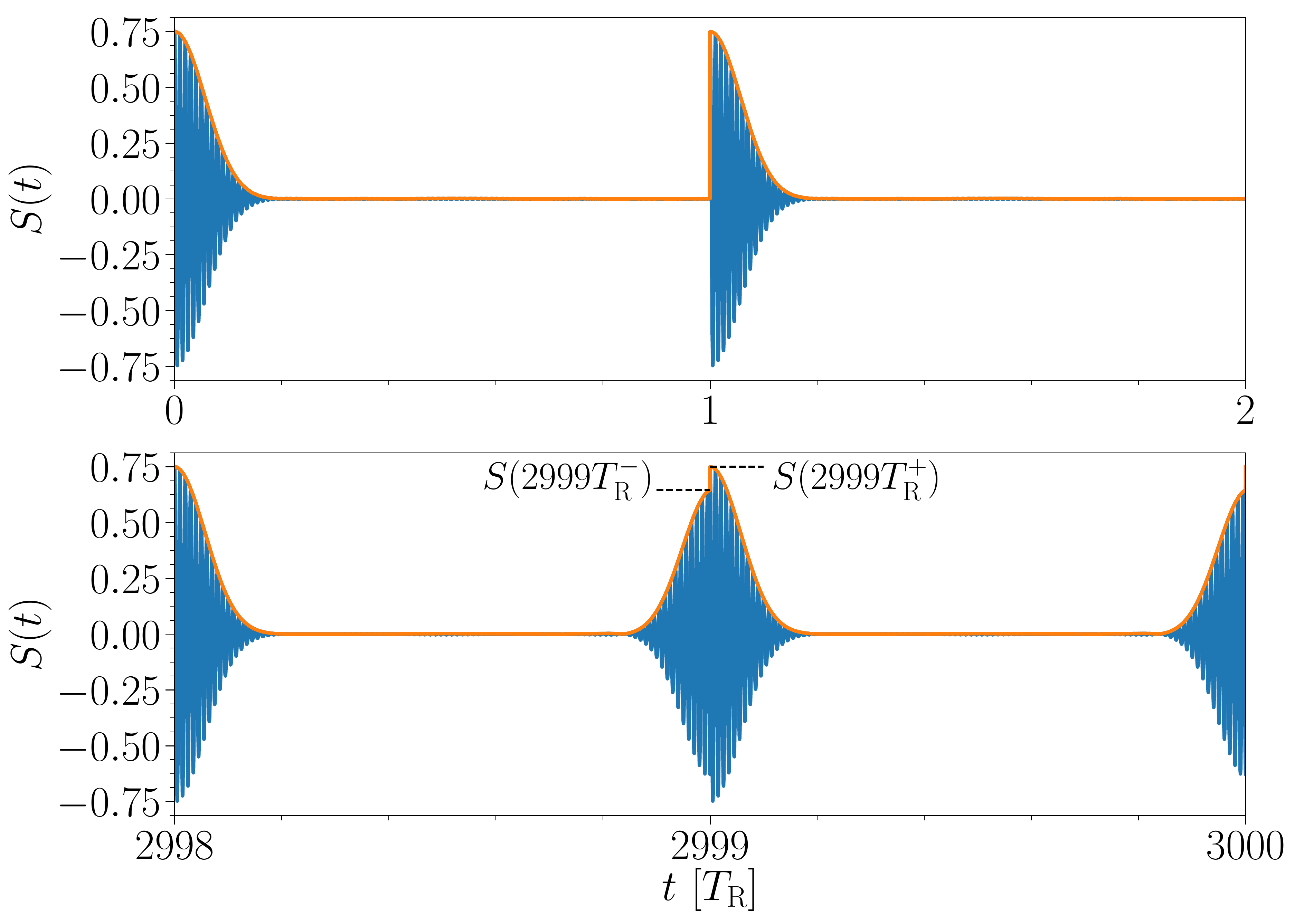}
\caption{
Representative result for the spin-spin correlation \eqref{eq:measure_def} at $h=40\jq$ and 
pulse repetition time $T\R=5\pi/\jq$ for a spin bath with $\gamma = 10^{-2}$ using pulse model I 
from \eqref{eq:pulse1}. The solid orange line indicates the envelope \eqref{eq:measure_def3}.}
\label{fig:representative_1}
\end{figure}

Figure \ref{fig:representative_1} displays a representative illustration of 
refocusing of the central spin precessions which occur after periodic pulsing with pulse \eqref{eq:pulse1} and 
repetition time $T\R = 5 \pi / \jq$, which is roughly in the experimentally relevant range
of $13.2$ns \cite{greil06a,greil06b,greil07a}.
We choose a multiple of $\pi/\jq$ in order to make commensurability effects easy to discern.
The origin of this refocusing is the nuclear frequency focusing which we will analyze below.

In order not to be distracted by the fast Larmor precessions, we
address the envelope directly by defining
\begin{subequations}
\begin{align}
\label{eq:measure_def2}
S^{yz}(t) &:= \overline{S_0^y(t)S_0^z(0)}
\\
\label{eq:measure_def3}
S(t) &:= \sqrt{(S^{zz}(t))^2 + (S^{yz}(t))^2}
\end{align}
\end{subequations}
and using \eqref{eq:measure_def}.
The modulus $S(t)$ represents the envelope of a fast precession about
the $S^x$-axis along which the external magnetic field is oriented. 
In Fig.\ \ref{fig:representative_1}, it is indicated by the solid orange line.

\begin{figure}[htb]
\centering
\includegraphics[width=1.0\columnwidth]{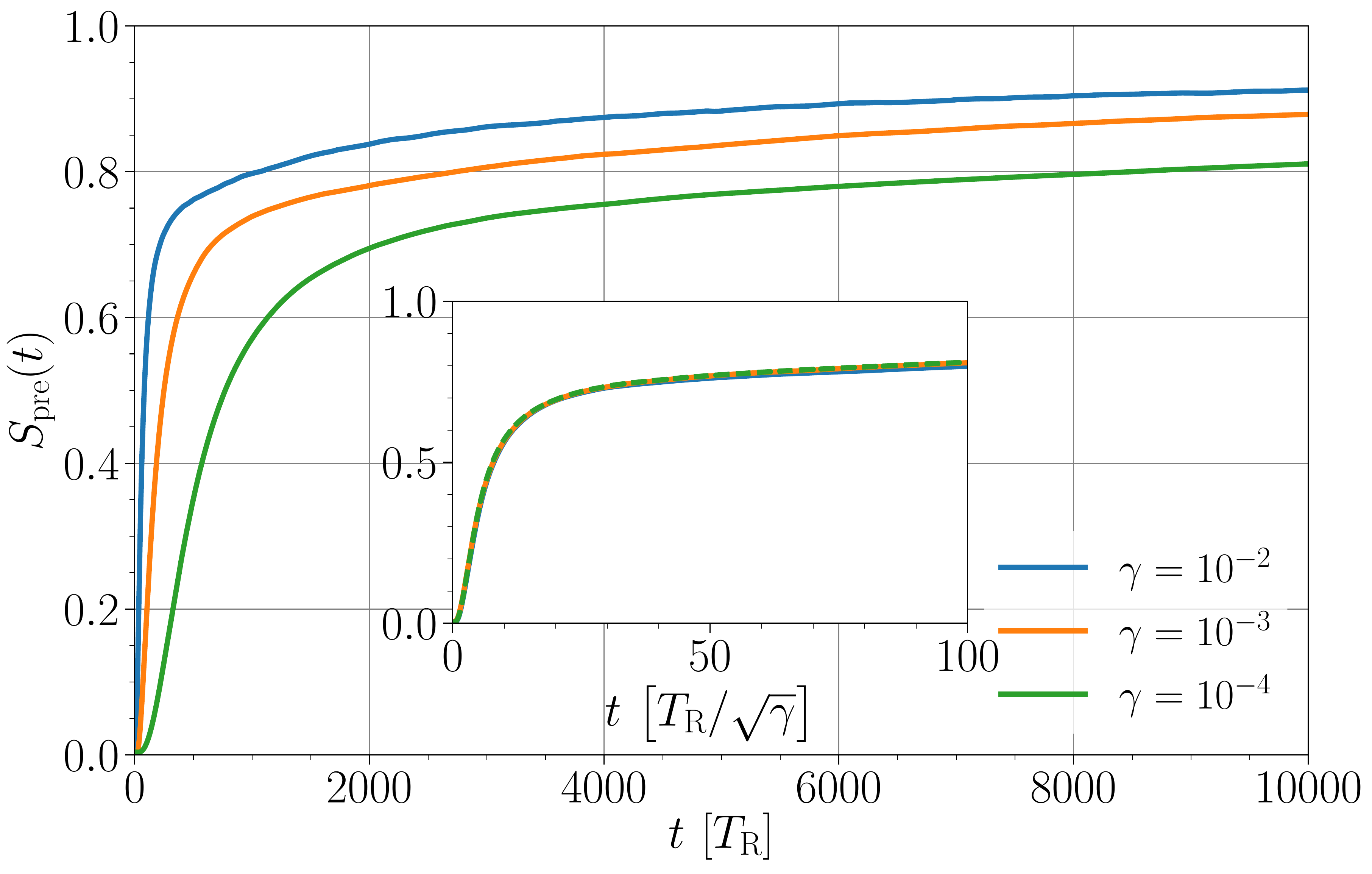}
\caption{
Relative pre-pulse signal as a function of time at $h=40\jq$ and pulse repetition time
$T\R=5\pi/\jq$ for various spin baths using pulse model I from \eqref{eq:pulse1}.
Inset: Excellent data collapse when the time axis is scaled with $\sqrt{\gamma}$.}
\label{fig:prepulse_1}
\end{figure}

The degree of nuclear focusing can be characterized by the 
relative pre-pulse amplitude
\begin{align}
S_\mathrm{pre}(t) := 
\frac{S(n T\R^{-})}{S(n T\R^{+})}, \qquad n \in \{1,2,\dots,n_\mathrm{p}\},
\end{align}
which is the quotient of the envelopes just before ($nT\R^-$) and after ($nT\R^+$)
a pulse is applied (see Fig.\ \ref{fig:representative_1} at $t/T\R = 2999$).
Figure \ref{fig:prepulse_1} shows this relative pre-pulse amplitude
for various values of the parameter $\gamma$.
Clearly, smaller values of $\gamma$ correspond to a larger effective spin $\neff$ bath, which leads to a slower build-up of the pre-pulse signal. 
This is expected because each individual coupling $J_k$ scales like $\sqrt{2\gamma}$
so that the dynamics of each bath spin is slower for decreasing $\gamma$.

In view of demanding numerical simulations, we like to make use of scaling arguments 
to reduce the amount of required computational resources.
In Ref.\ \onlinecite{fause17a}, a scaling of time with $\sqrt{\gamma}$ has been established 
for $h=0$ and without periodic pulsing.
The inset in Fig.\ \ref{fig:prepulse_1} shows that this scaling is still valid at a finite external field 
of $h=40 \jq$ for the pre-pulse amplitude $S_\mathrm{pre}(t)$ under 
periodic pulsing with pulse \eqref{eq:pulse1}.
This allows us to study the $h$-dependence later for only one particular value
of $\gamma = 10^{-2}$ and to scale the results up if needed for smaller values
of $\gamma$.

As mentioned above, the origin of the pre-pulse signal in the dynamics of the central spin in this fully classical simulation is the nuclear frequency focusing of the Overhauser field distribution along the axis of the external magnetic field, i.\,e., $B^x$ (Voigt geometry).
The basic idea is the following: each pulse kicks the electronic spin and 
its motion in turn has a small effect on each bath spin.
As long as the electronic spin does not precess with a Larmor frequency commensurate 
with the pulse repetition rate, these kicks continue to influence the state of the
bath spins. Only once commensurability is reached, the periodic pulses cease
to influence the distribution of bath spins: a quasi-stationary state
is reached. Thus, after long pulsing 
one expects a stationary distribution of the bath spins which is strongly peaked
at those values for the Overhauser field which induce commensurate precession
\cite{greil06b,greil07a,petro12,jasch17}.

In previous studies, the distribution of $B^x$ was investigated. But
our simulations showed that this is not sufficient to analyze the 
proper resonance conditions. For instance, it may even occur that 
even and odd resonance appear to be swapped, i.e., the integer and half-integer
number of spin revolutions between two pulses appear to be interchanged,
for details see below. Hence, we study the full effective Larmor frequency 
$\omega_\mathrm{L} = |\vec{h}_\mathrm{eff}| = |\vec{h} - \vec{B}|$, 
which is the relevant quantity. For strong external magnetic fields,
the distributions of $|\vec{h} - \vec{B}|$ and of $B^x$ are very similar.

\begin{figure}[htb]
	\centering
	\includegraphics[width=1.0\columnwidth]{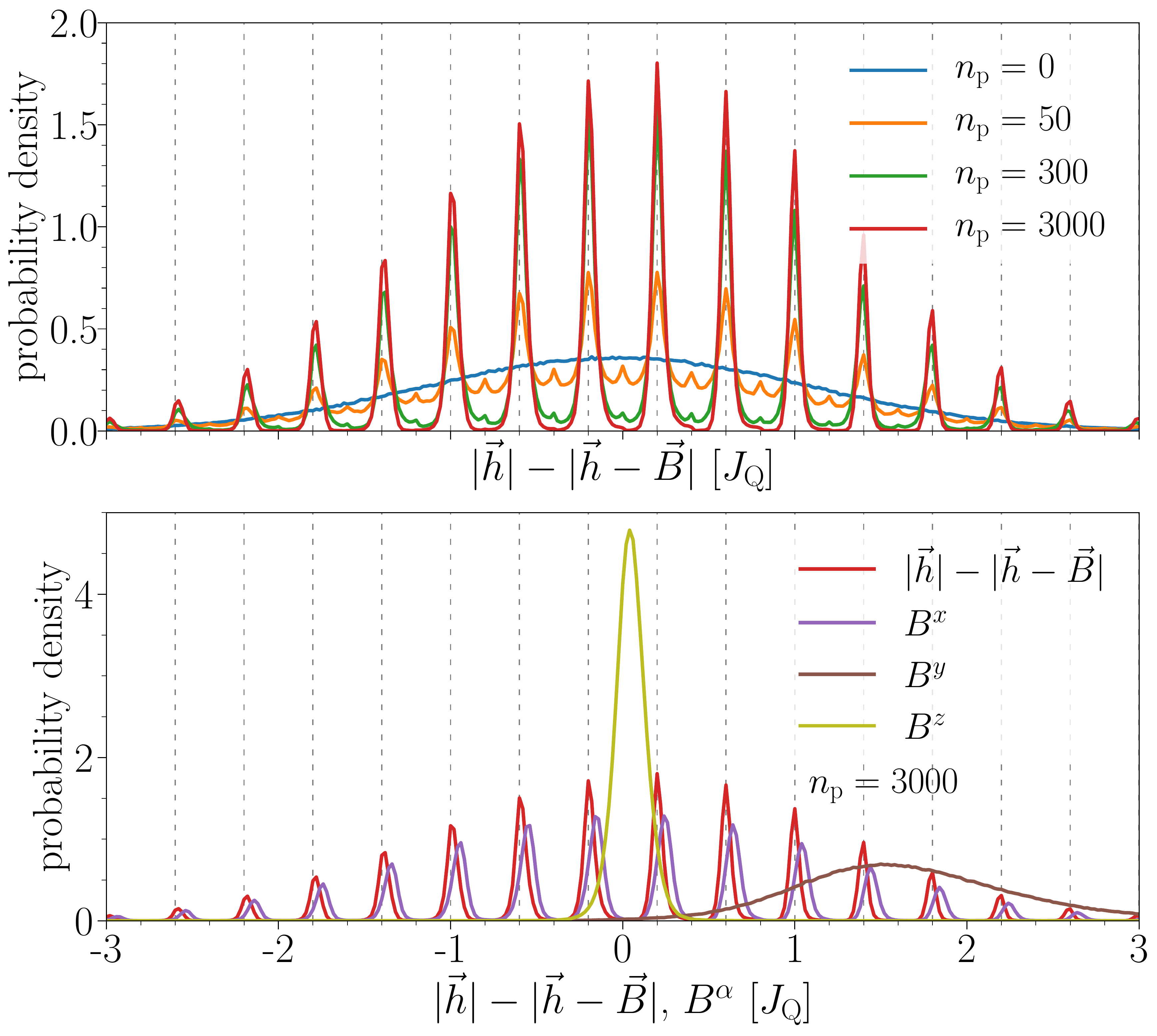}
	\caption{Upper panel: Distribution of the effective Larmor frequency $|\vec{h} - \vec{B}|$, shifted by $|\vec{h}|$, at $h=40\jq$ and pulse repetition time $T\R=5\pi/\jq$ 
	for $\gamma = 10^{-2}$ after $n_\mathrm{p}$ pulses using pulse model I from \eqref{eq:pulse1}. The vertical dashed lines indicate the 
	values which satisfy the odd resonance condition \eqref{eq:odd}.
	Lower panel: Same as upper panel, but after $n_\mathrm{p} = 3000$ pulses. Additionally, 
	the distributions of the Overhauser field components $B^\alpha$ ($\alpha\in\{x,y,z\}$) are shown.}
	\label{fig:Overhauser_1}
\end{figure}

The upper panel in Fig.\ \ref{fig:Overhauser_1} illustrates the dynamic build-up of 
nuclear frequency focusing after $n_\mathrm{p}$ pulses.
The distribution starts as a Gaussian with variances $\overline{(B^\alpha)^2} = 5\jq^2/4 $ ($\alpha\in\{x,y,z\}$)
as given by the initial condition \eqref{eq:overhaus_variance}.
Then, due to periodic pulsing, peaks are established over time. They appear to be perfectly centered around the values of $|\vec{h} - \vec{B}|$ which satisfy the odd resonance condition
\begin{align}
	|\vec{h} - \vec{B}|_\mathrm{odd} T\R = (2m + 1) \pi, \qquad m \in \mathbb{Z},
	\label{eq:odd}
\end{align}
indicated by the dashed vertical lines in Figure \ref{fig:Overhauser_1}.
This resonance condition is termed to be `odd' because $2m+1$ is an odd number. 
It corresponds to a half-integer number of {spin revolutions} of the central spin within the pulse interval $T\R$.
Small sub-peaks can be observed for the first few hundred pulses between the main peaks.
Their positions correspond to the even resonance condition
\begin{align}
	|\vec{h} - \vec{B}|_\mathrm{even} T\R = 2m \pi, \qquad m \in \mathbb{Z}\,,
	\label{eq:even}
\end{align}
with $2m$ being an even number, corresponding to an integer number of 
{spin revolutions} within the pulse interval $T\R$.
However, this mode appears to be suppressed quite fast, i.\,e., 
after a few hundred pulses.

How do the single components of the Overhauser field distribution $B^\alpha$ evolve?
An illustration is given {in the lower panel of}
 Fig.\ \ref{fig:Overhauser_1}, where the distributions of the $B^\alpha$ are shown together with {the one of} $|\vec{h} - \vec{B}|$ after $n_\mathrm{p} = 3000$ pulses.
Obviously, the comb-like structure of $|\vec{h} - \vec{B}|$ mainly stems from the distribution of $B^x$, i.\,e., along the external magnetic field direction.
This is in accordance with what was found in previous works \cite{greil06b,greil07a,petro12,jasch17}.
However, the peaks found for $B^x$ are shifted
 slightly away from the resonance condition \eqref{eq:odd}.
This is due to the finite contributions from $B^y$ and $B^z$, 
which also have an influence on the resonance according to \eqref{eq:odd}.
For larger magnetic fields, their {importance decreases rapidly}.

Interestingly, we find that $B^y$ is shifted to a finite mean value $\overline{ B^y } > 0$ and becomes slightly sharper.
For $B^z$, we observe a strong narrowing effect such that its contribution to the resonance almost vanishes.
The finding that the periodic pulsing can also generate non-trivial Overhauser fields perpendicular to the external magnetic field
and to the direction of the polarization of the pulses {carries} an interesting message also {for} experiment.
It would be interesting to devise measurements of the perpendicular Overhauser field, which we find to be of the order of
$\overline{B^y} \approx 100\,\mathrm{mT}$ after {long} pulsing. 
To our knowledge, however, such effects have not yet been observed experimentally.

\begin{figure}[htb]
	\centering
	\includegraphics[width=1.0\columnwidth]{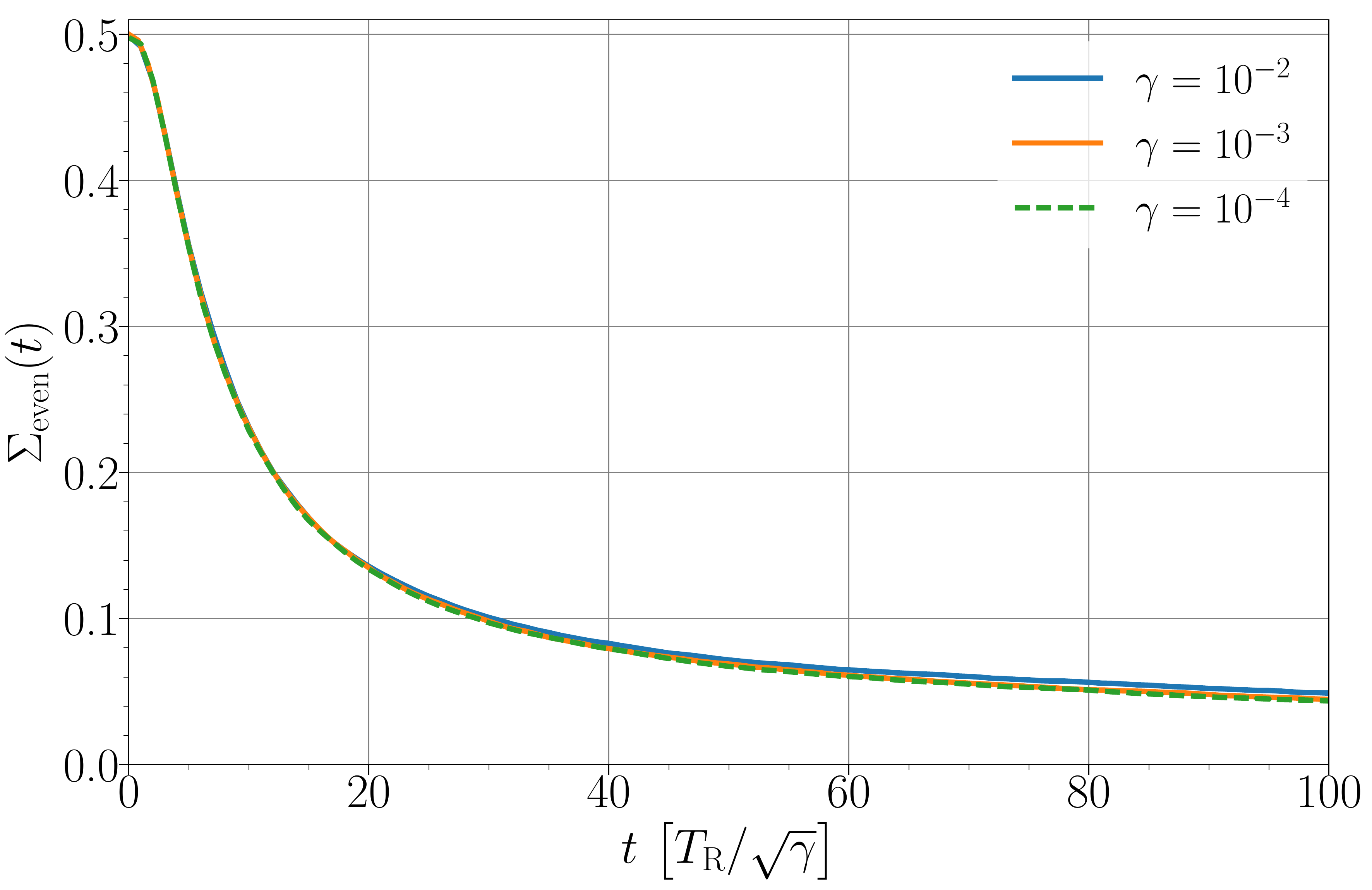}
	\caption{Weight $\Sigma_\mathrm{even}$ of the Overhauser field distribution fulfilling 
	the even resonance condition \eqref{eq:even} as a function of the scaled time 
	$t\sqrt{\gamma}$ at $h=40\jq$ and pulse repetition time $T\R=5\pi/\jq$ for various 
	spin baths using pulse model I from \eqref{eq:pulse1}.}
	\label{fig:scaled_weight_gamma_1}
\end{figure}

In order to study the build-up process quantitatively, 
we analyze the weight of the even and odd {resonance} by 
defining the following bins.
The even {resonance} is characterized by all values $|\vec{h}-\vec{B}|$ in the 
bins $\left[|\vec{h}-\vec{B}|_\mathrm{even} - \frac{\pi}{2T\R},\, 
|\vec{h}-\vec{B}|_\mathrm{even} + \frac{\pi}{2T\R} \right]$. 
Their relative number defines the weight $\Sigma_\mathrm{even}$.
Analogously, the odd {resonance} is characterized by all values 
$|\vec{h}-\vec{B}|$ in the bins 
$\left[|\vec{h}-\vec{B}|_\mathrm{odd} - \frac{\pi}{2T\R},\, 
|\vec{h}-\vec{B}|_\mathrm{odd} + \frac{\pi}{2T\R} \right]$. 
Obviously, the relation
\begin{align}
	\Sigma_\mathrm{even}(t) + \Sigma_\mathrm{odd}(t) = 1
\end{align}
holds, hence it is sufficient to investigate only 
one of the two weights. We choose $\Sigma_\mathrm{even}$.
In case of perfect even resonance, this weight rises to unity.
 In case of perfect odd resonance, this weight shrinks to zero.
Figure \ref{fig:scaled_weight_gamma_1} depicts
$\Sigma_\mathrm{even}(t)$ for various values of $\gamma$ and with scaled time $t$
displaying a clear signature of odd resonance.
Again, the scaling with $\sqrt{\gamma}$ leads to a remarkably good collapse
of the curves. 
Therefore, we conclude that the degree of nuclear frequency focusing in the Overhauser field,
i.\,e., the convergence of $\Sigma_\mathrm{even}$ to zero, influences directly
the relative strength of the pre-pulse signal of the central spin.

\begin{figure}[htb]
	\centering
	\includegraphics[width=1.0\columnwidth]{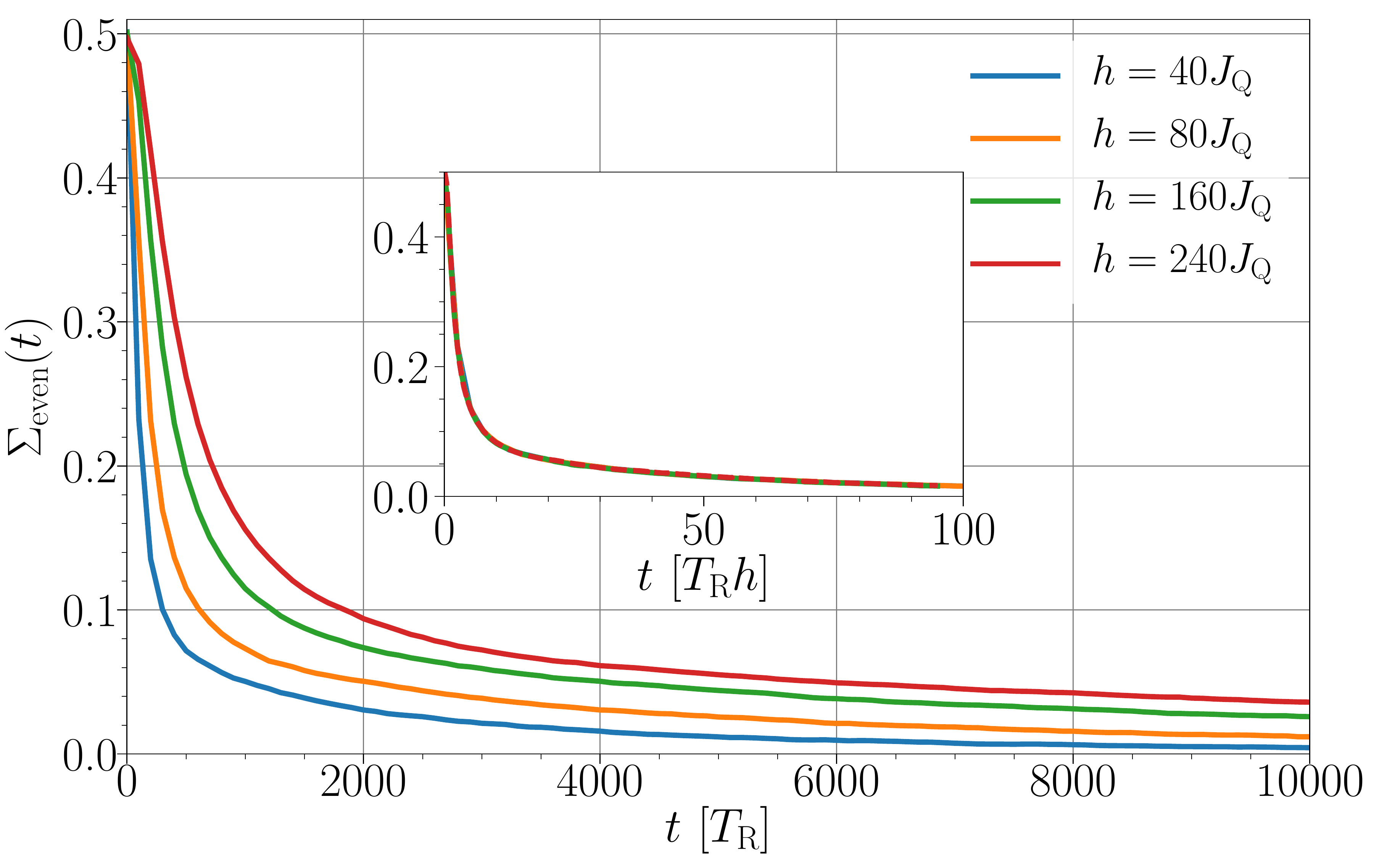}
	\caption{Weight $\Sigma_\mathrm{even}$ of the Overhauser field distribution fulfilling the
	even resonance condition \eqref{eq:even} as a function of time for $\gamma = 10^{-2}$ and
	pulse repetition time $T\R=5\pi/\jq$ for various external magnetic fields $h$ using pulse model I from \eqref{eq:pulse1}.
	Inset: Excellent data collapse when the time axis is scaled with $1/h$.}
	\label{fig:weight_h_1}
\end{figure}

Up to now we only studied rather small external fields, i.\,e., $h = 40 \jq$.
The question arises as to what happens when $h$ is increased.
The experimental values are range from $1\,\mathrm{T}$ to $6\,\mathrm{T}$, corresponding to field strengths $h \in [40 \jq,\, 240 \jq]$ in our units.
Figure \ref{fig:weight_h_1} shows the time dependence of the weight 
$\Sigma_\mathrm{even}$ for various values of $h$.
In general, a larger magnetic field corresponds to a slower build-up of nuclear focusing
\cite{beuge16}.
We find that a scaling of time with $1/h$ leads to a perfect collapse of the curves as shown in the inset of Fig.\ \ref{fig:weight_h_1}.

The linear scaling at high magnetic fields is at odds with the quantum mechanical analysis
\cite{beuge16} which indicates quadratic scaling $\propto 1/h^2$. The quadratic scaling
has been found in a semi-classical analysis as well \cite{glazov12}, 
but using a different pulse model and including the nuclear Zeeman term.
Indeed, we suppose that the classical analysis using pulse model I differs 
in this respect from the quantum mechanical behavior, see also next section.

Another way to investigate the resonance behavior is to compare the phases
of the pre-pulse signal to the post-pulse signal. If there is a phase jump by $\pi$,
the resonance is odd. If there is no phase jump, the resonance is even.
In principle, also other phase jumps could occur. 
The details of this analysis are discussed in Appendix \ref{App:phase}.
We always find a phase jump of $\pi$ within 2 to 3\% for the pulse model I 
studied in this section{, in agreement with the analysis of 
$\Sigma_\mathrm{even}\to 0$.}

Thus, this pulse {model} clearly favors odd resonances in agreement with what is 
found without nuclear Zeeman effect in a quantum mechanical analysis for small
spin baths \cite{beuge16}. The nuclear Zeeman effect has
a pronounced effect on the resonance condition \cite{beuge17,jasch17}.
It shifts the resonances from odd to even.  We come back to this point
in subsequent sections. If the experimental situation of
various different nuclear $g$-factors is considered the phenomenology becomes
even richer \cite{beuge17}.

We summarize that periodic pulsing with pulse model I shows
efficient nuclear frequency focusing at odd resonance.
The rate of change of the Overhauser field scales with $\sqrt{\gamma}$,
i.\,e., inversely proportional to the square root of the effective size of the spin bath.
In addition, it scales inversely proportional to the magnetic field which
is at odds with the finding of a quantum mechanical calculation \cite{beuge16}.
We attribute this discrepancy to the classical nature of pulse model I.
The scalings in $\sqrt{\gamma}$ and $1/h$ generate very nice data collapses so
that quantitative extrapolations are possible with a high degree of accuracy.

\subsection{Pulse model II from Eq. \eqref{eq:pulse2}}
\label{subsec:NNZ_2}

Next, we turn to pulse model II and carry out the same analyses as for pulse model I. 
The motivation is twofold. First, we want to see to
which extent the previous findings change if the pulse is changed. The underlying
issue is whether and to which extent the findings are robust to
the details of the pulse. Second, we assume that pulse model II is closer to
a quantum mechanical pulse and hence we are interested in its
phenomenology.

\begin{figure}[htb]
	\centering
	\includegraphics[width=1.0\columnwidth]{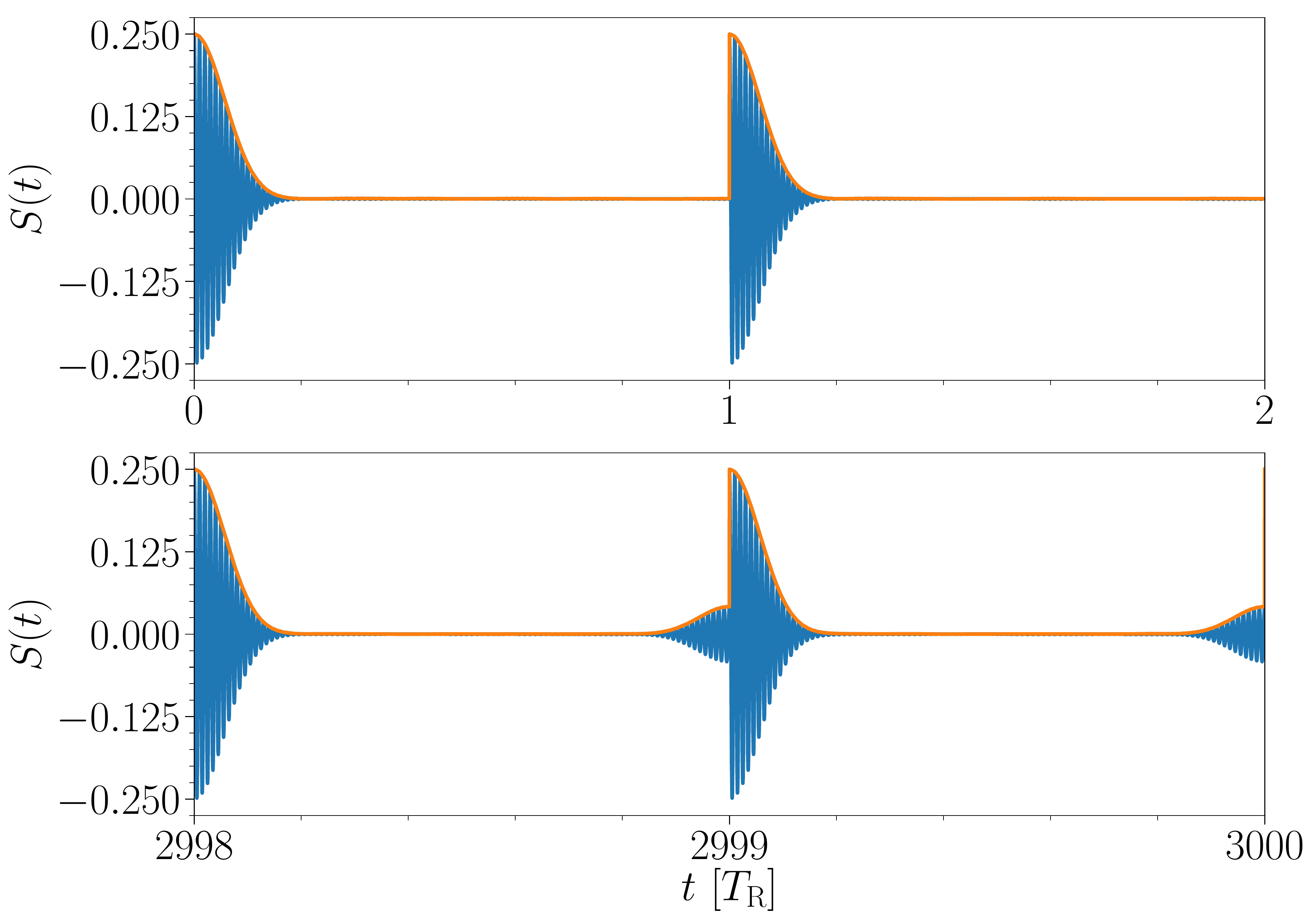}
	\caption{
		Representative result for the spin-spin correlation \eqref{eq:measure_def} at $h=40\jq$ and 
pulse repetition time $T\R=5\pi/\jq$ for a spin bath with $\gamma = 10^{-2}$ using pulse model II
from \eqref{eq:pulse2}. The solid orange line indicates the envelope \eqref{eq:measure_def3}.}
	\label{fig:representative_2}
\end{figure}

\begin{figure}[htb]
\centering
\includegraphics[width=1.0\columnwidth]{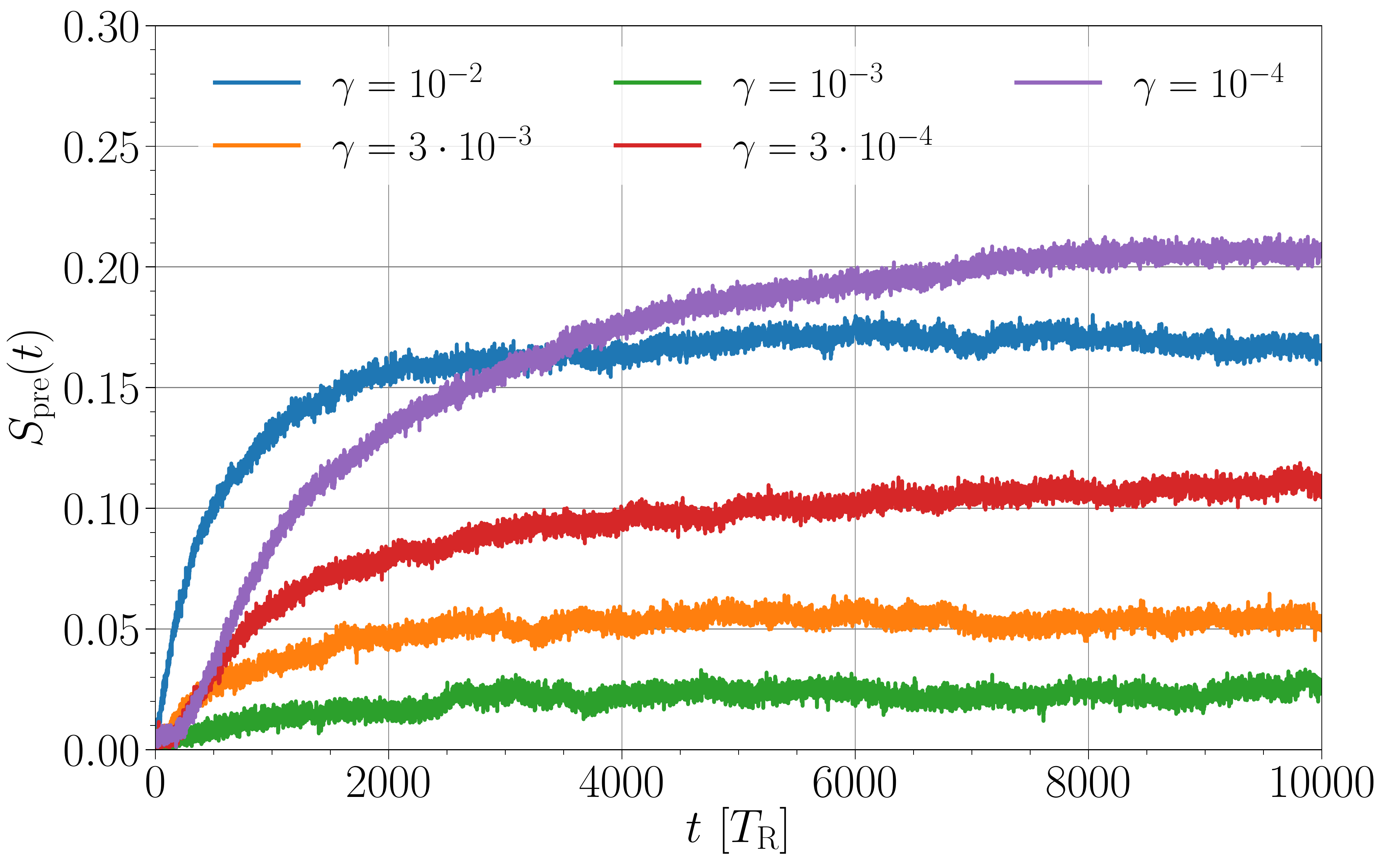}
\caption{
	Relative pre-pulse signal as a function of time at $h=40\jq$ and pulse repetition time
	$T\R=5\pi/\jq$ for various spin baths using pulse model II from \eqref{eq:pulse2}.}
\label{fig:prepulse_2}
\end{figure}

Indeed, we find qualitative differences.
The tendency to nuclear focusing is much less pronounced 
for pulse model II than for pulse model I. This can be seen by
comparing the lower panel of Fig.\ \ref{fig:representative_2} to the lower panel in
Fig.\ \ref{fig:representative_1}. The comparison of Fig.\ \ref{fig:prepulse_2}
showing the relative pre-pulse signal
to Fig.\ \ref{fig:prepulse_1} is more quantitative. 
The obvious feature
is that the pre-pulse signal of pulse model II does not reach the high values
of the one of pulse model I (note the different scale on the $y$-axis). 
In addition, the data is noisier being characterized by more fluctuations.

The most striking feature, however, is the \emph{non-monotonic} behavior on the 
effective spin bath size $\neff=2/\gamma$. The strongest increase is
found for $\gamma=10^{-2}$, but the increase of the relative pre-pulse
signal for $\gamma=10^{-4}$ is slower only by about a factor of 2
and reaches even \emph{higher} saturation values. Obviously, no
scaling with any power of $\gamma$ will lead to a collapse
of curves. Quite unexpectedly,
the curve for $\gamma=10^{-3}$ increases slower than the other two curves
and does not reach a significant value at all. 
The curves for $\gamma = 3 \cdot 10^{-3}$ and $3 \cdot 10^{-4}$ are still showing a rather weak pre-pulse signal, but stronger than for $\gamma = 10^{-3}$. 
It appears that there is {a qualitative} transition occurring at 
around $\gamma = 10^{-3}$. We come back to this point below.

In addition, we stress that
the saturation values appear to stay far away from the theoretical
maximum of unity, i.\,e., the periodic pulsing with pulse model II induces
only non-perfect nuclear focusing. The contrast of these results to the
ones for pulse model I underlines the importance of the pulse properties.

\begin{figure}[htb]
	\centering
	\includegraphics[width=1.0\columnwidth]{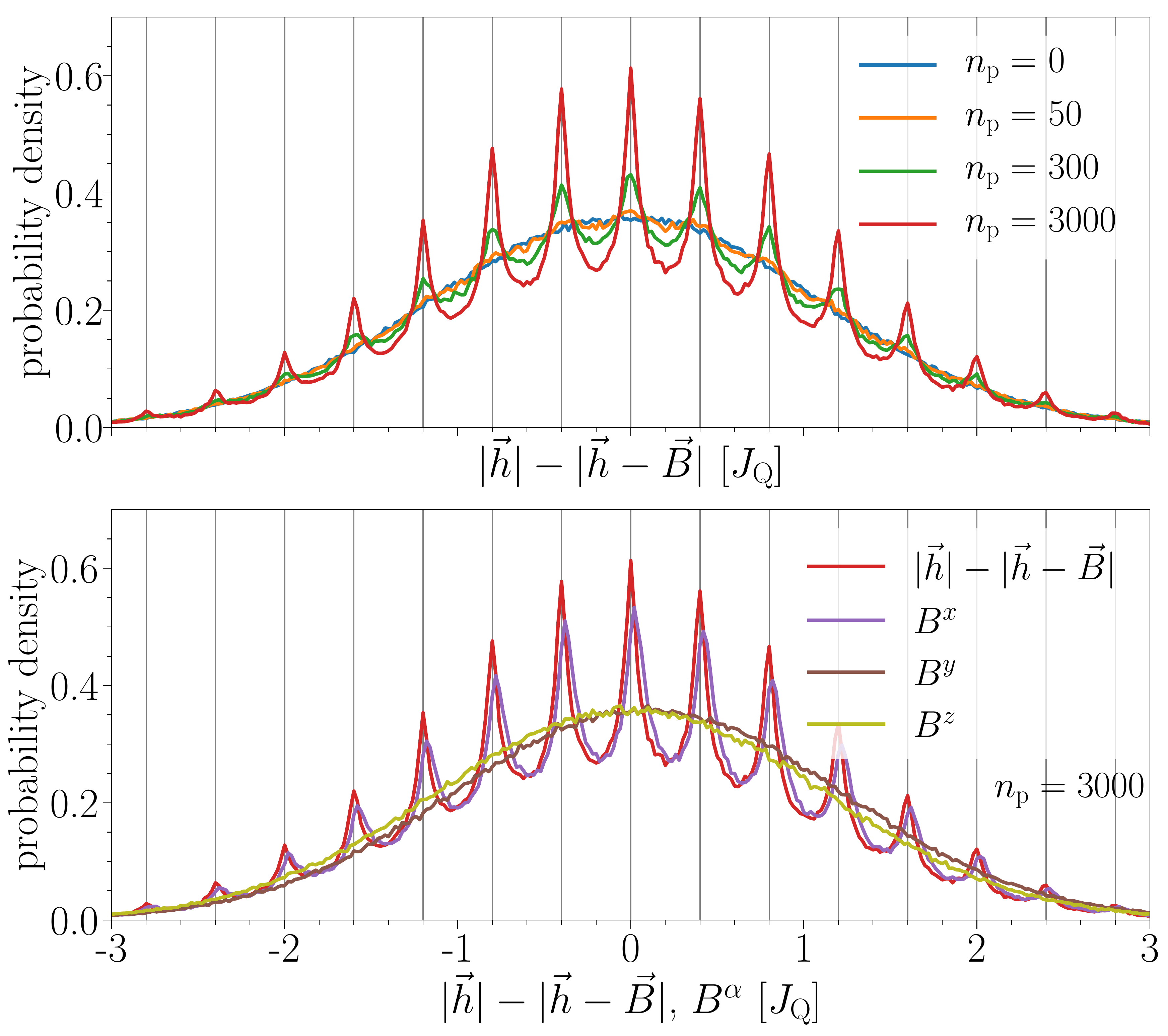}
	\caption{Upper panel: Distribution of the effective Larmor frequency $|\vec{h} - \vec{B}|$, shifted by $|\vec{h}|$, at $h=40\jq$ and pulse repetition time $T\R=5\pi/\jq$ 
	for $\gamma = 10^{-2}$ after $n_\mathrm{p}$ pulses using pulse model II from \eqref{eq:pulse2}. The vertical solid lines indicate the 
	values which satisfy the even resonance condition \eqref{eq:even}.
	Lower panel: Same as upper panel, but after $n_\mathrm{p} = 3000$ pulses. Additionally, 
	the distributions of the Overhauser field components $B^\alpha$ ($\alpha\in\{x,y,z\}$) are shown.}
	\label{fig:Overhauser_2}
\end{figure}

How does the Overhauser field evolve if driven by periodic pulses of type II?
The evolution of a representative distribution of the effective Larmor frequency is
depicted in the upper panel of Fig.\ \ref{fig:Overhauser_2} for 
$h = 40 \jq$ and $\gamma = 10^{-2}$.
The peaks of the effective Larmor frequency $|\vec{h} - \vec{B}|$ 
{are much broader and lower} and therefore the degree of 
nuclear focusing is much less pronounced 
compared to the distribution induced by pulse model I (see Fig.\ \ref{fig:Overhauser_1}).
The peaks in Fig.\ \ref{fig:Overhauser_2}
{are not located} at the magnetic field values corresponding to the odd
resonances, but they correspond to even resonances. 
Note that in contrast to the distribution obtained by applying pulse model I, the $y$- and $z$-components
of the Overhauser field maintain their initial Gaussian shape (lower panel).
Yet, the peaks found in the $B^x$ distribution are still slightly shifted to the right of the
theoretically expected resonance condition.
This is not the case for the effective Larmor frequency because it also incorporates
the variance of $B^y$ and $B^z$.

For long enough pulsing the distribution of the effective Larmor frequency $|\vec{h} - \vec{B}|$
becomes quasi-stationary. It does not change anymore
if it is analyzed stroboscopically, i.\,e., at a given instant relative
to the pulses, for instance just before each pulse.
But the distribution does not approach sharp peaks.
The peaks still keep an appreciable width and there is always some weight
around both resonance values, even and odd.
This qualitative behavior is confirmed quantitatively 
by the weight $\Sigma_\mathrm{even}(t)$ which approaches neither zero nor one.

\begin{figure}[htb]
	\centering
	\includegraphics[width=1.0\columnwidth]{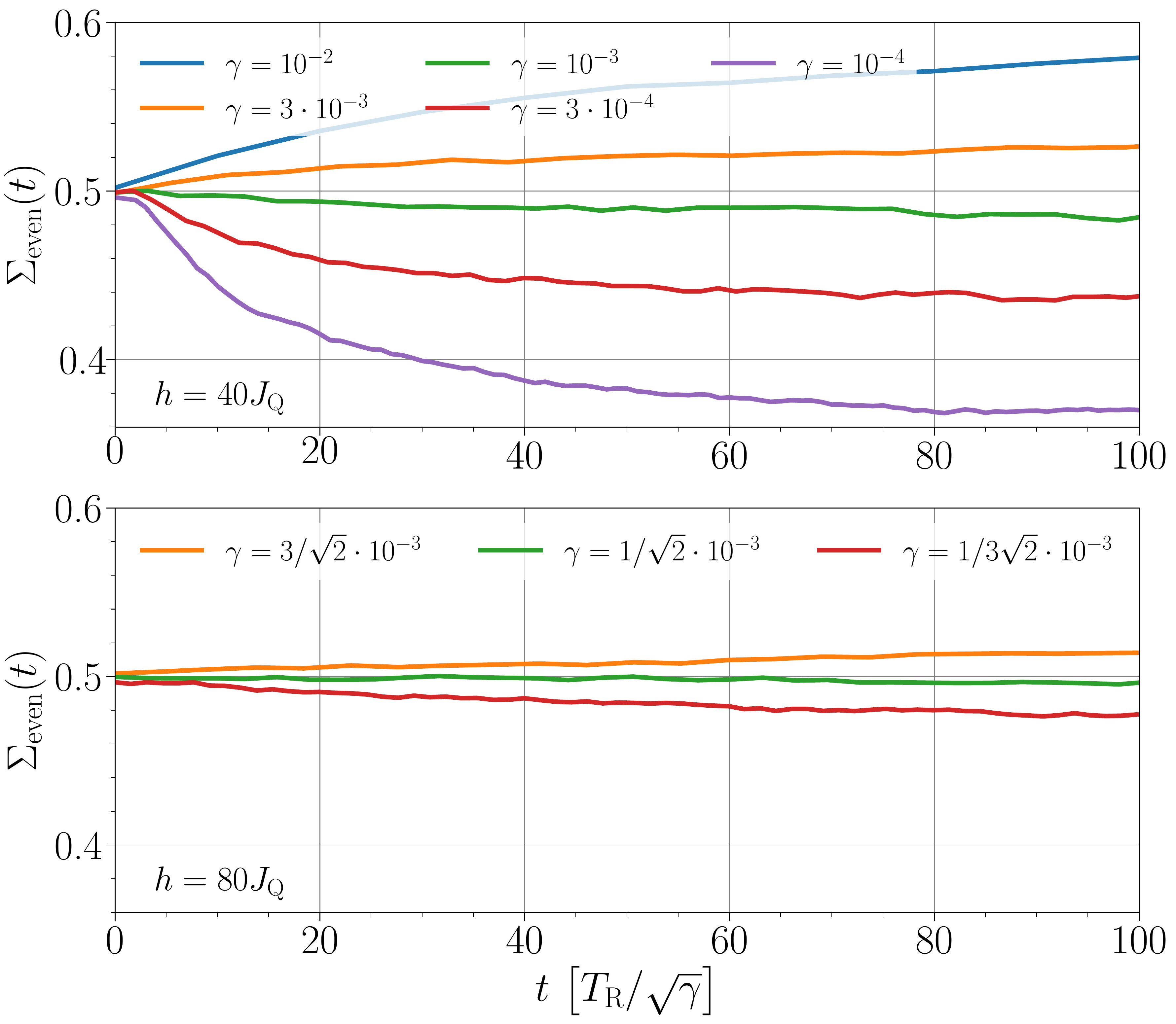}
	\caption{Upper panel: Weight $\Sigma_\mathrm{even}$ attributed to 
	the even resonance \eqref{eq:even} in the 
	Overhauser field distribution as a function of the scaled time $t\sqrt{\gamma}$ 
	at $h=40\jq$ and pulse repetition time $T\R=5\pi/\jq$ for various sizes of the spin baths 
	using pulse model II from \eqref{eq:pulse2}.
	Lower panel: Same as upper panel, but for $h=80\jq$ and adjusted sizes
	of the spin bath (see Eq.\ \eqref{eq:hgamma2}).}
	\label{fig:scaled_weight_gamma_2}
\end{figure}

Remarkably, there is a transition from even to odd resonance upon varying $\gamma$. 
This is shown in Fig.\ \ref{fig:scaled_weight_gamma_2} for the weight 
$\Sigma_\mathrm{even}$ at $h=40 \jq$ and $80 \jq$.
Upon reducing $\gamma$, the even resonance is replaced by the odd resonance.
This does not happen by a continuous shift of the peak positions in the Overhauser field distribution, but the distribution becomes featureless at the transition
between even and odd resonance, for instance at
 $\gamma \approx 10^{-3}$ and $h=40\jq$ {(not shown)}.
We do not observe coexistence of even and odd resonance peaks
as was found in the semi-classical analysis in Ref.\ \onlinecite{jasch17}.

We emphasize that Fig.\ \ref{fig:scaled_weight_gamma_2}
 indicates that the change of the Overhauser field
distribution happens at a rate proportional to $\sqrt{\gamma}$. Even though we
change $\gamma$ by two orders of magnitude the typical slopes occurring
in Fig.\ \ref{fig:scaled_weight_gamma_2} are the same. The impossibility to
achieve a data collapse by scaling the time is explained 
by the non-monotonic behavior of $\Sigma_\mathrm{even}$ as a function
of $\gamma$ due to the transition between even and odd resonance.
But the generic time constant remains $\sqrt{\gamma} \jq$ as for pulse model I.

To corroborate the {existence of the}
transition between even and odd resonance, we investigate
the phase jump $\Delta \varphi$  obtained by fitting the spin-spin correlation
just before and just after a pulse after long pulsing.
We analyze the combinations of $h$ and $\gamma$ used in Fig.\ 
\ref{fig:scaled_weight_gamma_2} and some other parameter combinations.
As expected, the phase jumps are directly connected to the parity of the resonance,
i.\,e., if $\Sigma_\mathrm{even} > 0.5$ no phase jump occurs while it takes the
value $\pi$ if $\Sigma_\mathrm{even} < 0.5$.
More details of this analysis are given in Appendix \ref{App:phase}.

This kind of transition has not yet been observed in other
calculations for the CSM. At present, we do not have an explanation
for its occurrence because there are so many energy scales in the problem
so that various combinations can become relevant. But we strive to 
provide a heuristic description for which parameters $h$ and $\gamma$ the transition occurs.
From a wide range of numerical experiments the working hypothesis ensues
that the transition occurs for 
\begin{align}
P := h\gamma^2 = \mathrm{const}.
\label{eq:hgamma2}
\end{align}

\begin{figure}[htb]
	\centering
	\includegraphics[width=1.0\columnwidth]{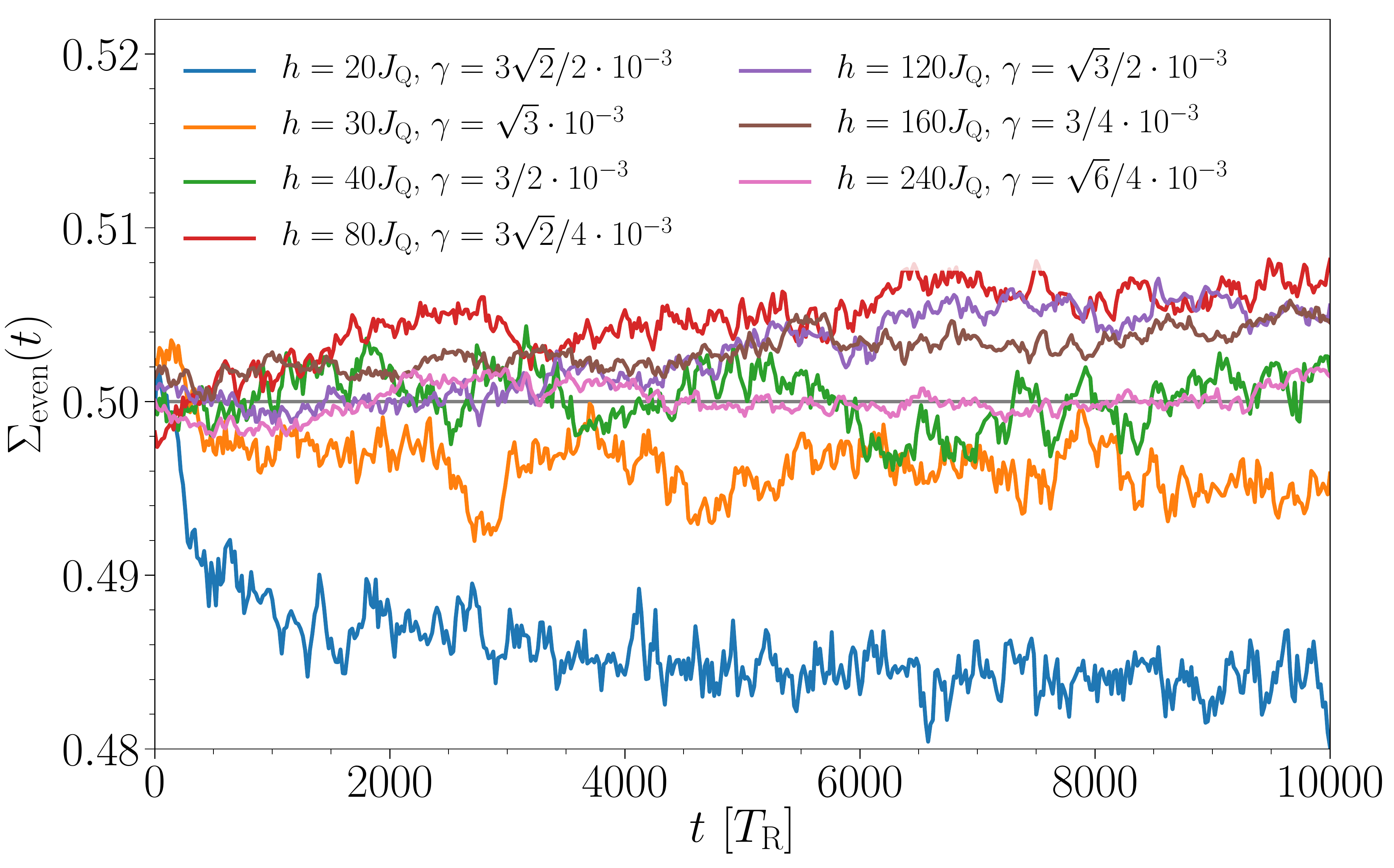}
	\caption{Various pairs of magnetic field $h$ and inverse spin bath size $\gamma$
	fulfilling $P=h\gamma^2 = 9 \cdot 10^{-5} \jq$ to test the conjecture \eqref{eq:hgamma2}. }
	\label{fig:hgamma2_const}
\end{figure}

We find that for  $h = 40\jq$ and $\gamma = 1.5 \cdot 10^{-3}$, no tendency towards even or odd resonance occurs
so that this parameter combination provides a valid estimate for $P$ yielding $P = 9 \cdot 10^{-5} \jq$. 
From this value, we generate other possible combinations  of $h$ and $\gamma$ which should correspond to parameters at the transition
according to the conjecture \eqref{eq:hgamma2}.
Figure \ref{fig:hgamma2_const} puts this conjecture to a test. 
Note the scale on its $y$-axis which is a factor $5$ {smaller} than in 
Fig.\ \ref{fig:scaled_weight_gamma_2}. Indeed, for large enough
magnetic fields the conjecture \eqref{eq:hgamma2} appears to hold within numerical
accuracy.
Note that this result is not very sensitive to the exact value of $P$. For instance, we were able to produce a similar plot as in Fig.\ \ref{fig:hgamma2_const} for $P = 3.8 \cdot 10^{-5}$ (not shown).

The issue of scaling the dynamics with the magnetic field $h$ arises.
Close to the transition we cannot expect a simple power-law scaling, just
as we did not find a power law scaling with $\gamma$ due to the transition.
Yet far away from the transition, the typical rate of change of the
Overhauser field distribution can be investigated to see whether a scaling
can be identified.

There are two ways to keep away from the transition. Either one stays
far in the regime of even resonance, i.\,e., for relatively small spin baths (large
values of $\gamma$)
at given magnetic field, or one stays far in the regime of odd resonance, i.\,e., for 
relatively large spin baths (small values of $\gamma$). If we take the above
determined value of $P= 9 \cdot 10^{-5} \jq$ and insert $\gamma=10^{-5}$
into \eqref{eq:hgamma2} we obtain $h=9 \cdot 10^5\jq$ corresponding to magnetic
fields of more than $2 \cdot 10^4\,\mathrm{T}$. Hence, quantum dots are expected to be in the
regime of odd resonance without nuclear Zeeman effect.
Other systems with much smaller spin baths, for instance NV centers or
spin in organic molecules, can very well be in the regime of even resonance.

\begin{figure}[htb]
	\centering
	\includegraphics[width=1.0\columnwidth]{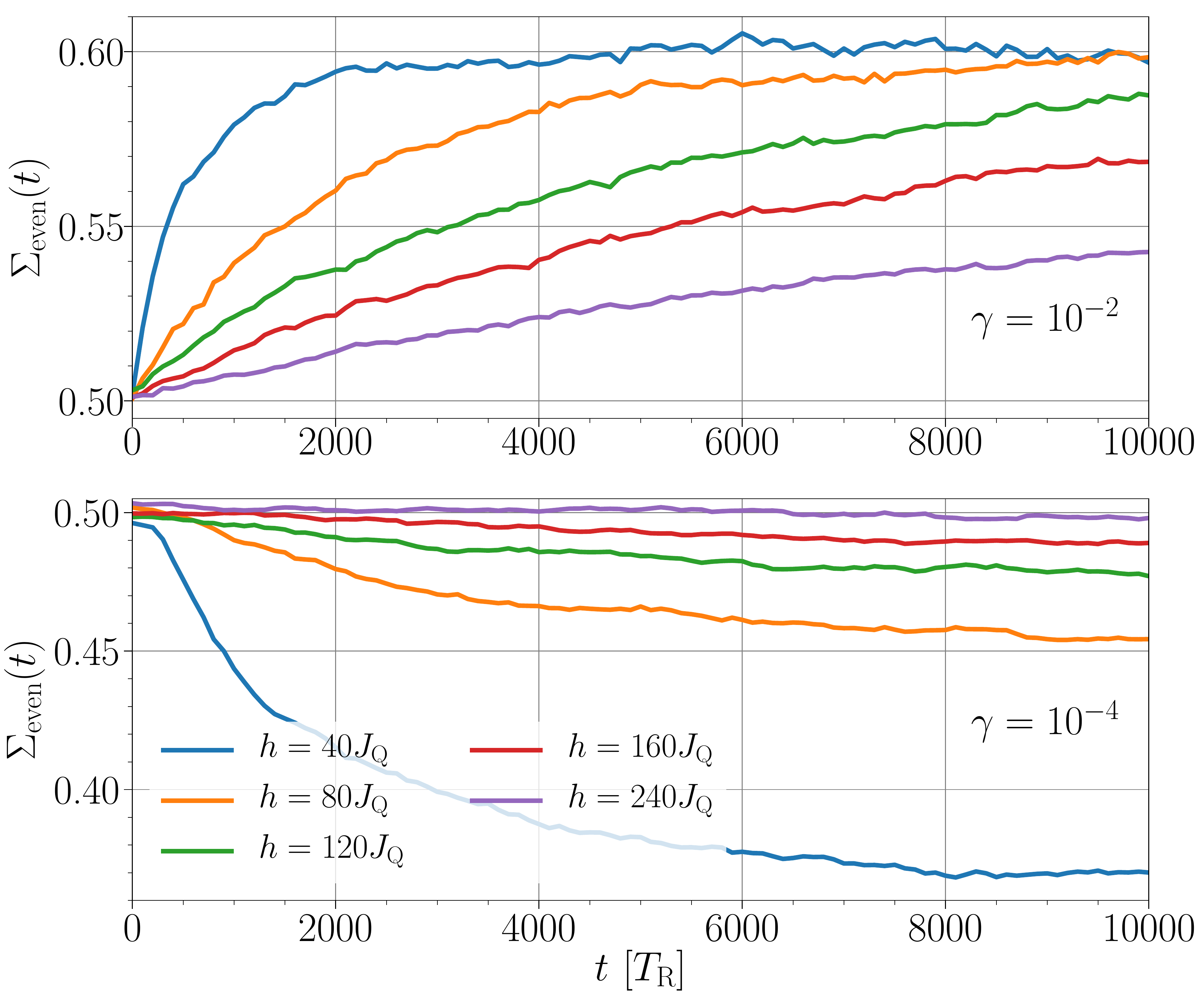}
	\caption{Upper panel: Weight $\Sigma_\mathrm{even}$ of the Overhauser fields generating even resonance  as a function of time for $\gamma = 10^{-2}$ and pulse repetition time $T\R=5\pi/\jq$ for various external magnetic fields $h$ using pulse model II from \eqref{eq:pulse2}.
	Lower panel: Same as upper panel, but for $\gamma = 10^{-4}$.}
	\label{fig:weight_h_2}
\end{figure}

\begin{figure}[htb]
	\centering
	\includegraphics[width=1.0\columnwidth]{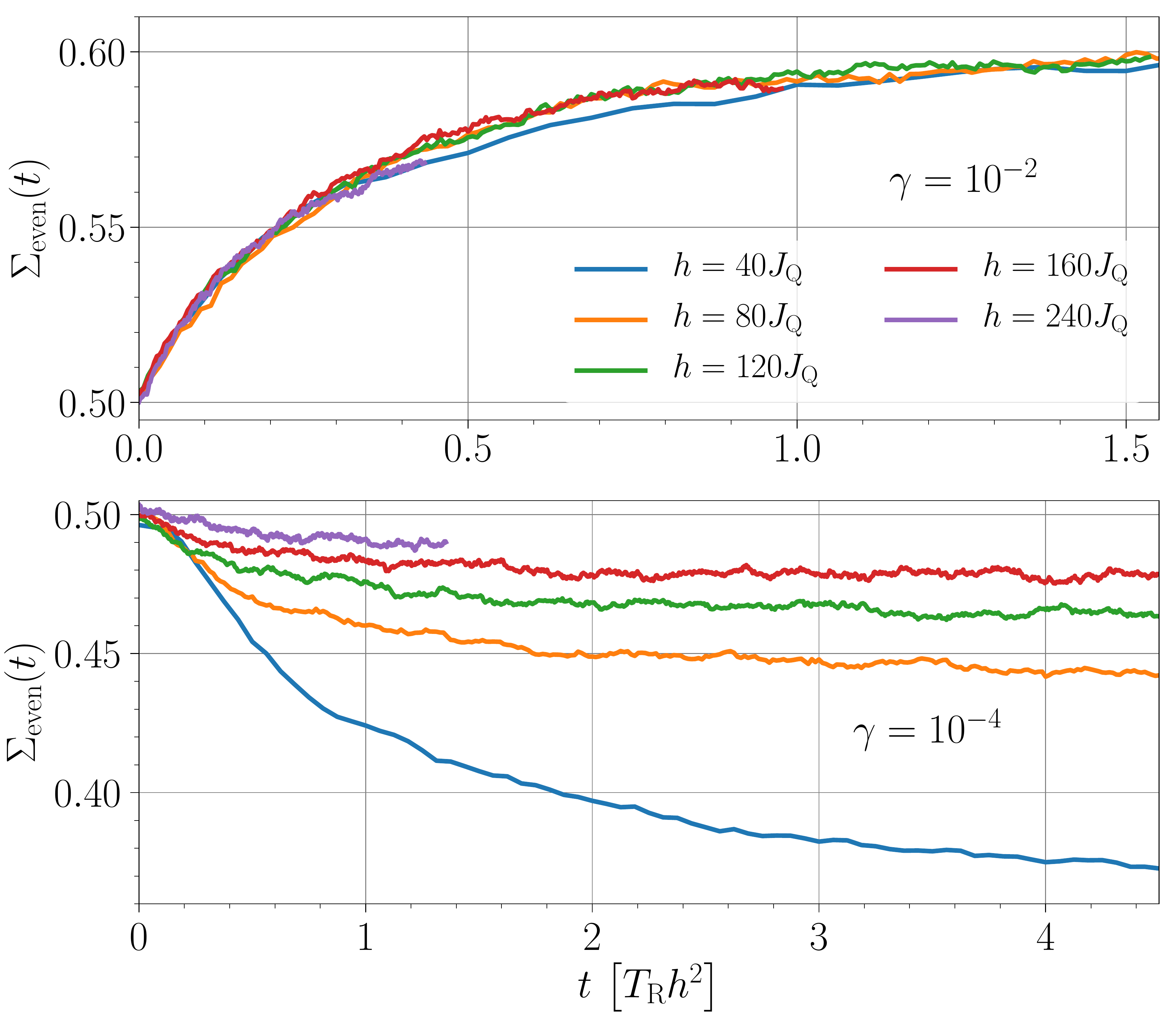}
	\caption{Weight $\Sigma_\mathrm{even}$ of the Overhauser fields generating even resonance  as a function of the scaled time $t/h^2$ for $\gamma = 10^{-2}$ and pulse repetition time 
	$T\R=5\pi/\jq$ for various external magnetic fields $h$ using pulse model II from \eqref{eq:pulse2}.
		Lower panel: Same as upper panel, but for $\gamma = 10^{-4}$.}
	\label{fig:scaled_weight_h_2}
\end{figure}

First, we study the regime of even resonance. The upper panel of Fig.\ \ref{fig:weight_h_2}
displays the increase of $\Sigma_\mathrm{even}$ for $\gamma=10^{-2}$ and various
magnetic fields. Clearly, larger magnetic field $h$ implies a slower build-up as expected.
Can we reach a data collapse by rescaling time by a power of $h$? 
For pulse model I, we succeeded in doing so with a linear scaling. This does
not hold here. Instead, the rate of the increase of $\Sigma_\mathrm{even}$
scales with $1/h^2$. This is shown by the collapse of curves
in the scaled plot rendered in the upper panel of Fig.\  \ref{fig:scaled_weight_h_2}.

The data collapse works nicely for the build-up of $\Sigma_\mathrm{even}$ (upper panel)
which agrees with the quantum mechanical \cite{beuge16} and the semi-classical result \cite{glazov12}.

Next, we address the regime of odd resonance
occurring for very large spin baths. The lower panel of Fig.\ \ref{fig:weight_h_2}
displays the decrease of $\Sigma_\mathrm{even}$ for $\gamma=10^{-4}$ and various
magnetic fields.
The rate at which $\Sigma_\mathrm{even}$ decreases scales approximately with $1/h^2$ as shown in the lower panel of Fig.\ \ref{fig:scaled_weight_h_2}.
However, the saturation values differ vastly for different magnetic fields $h$, with larger $h$ corresponding to a less pronounced odd resonance.

We summarize that pulse model II without nuclear Zeeman effect 
displays regimes of even and odd resonance with a 
transition between them depending on
the precise parameters of magnetic field and spin bath size.
According to the heuristic description of the position of the
transition by \eqref{eq:hgamma2}, the experimental setups for quantum
dots are far in the regime of odd resonance.
The overall rate of change of the Overhauser field scales proportional
to $\sqrt{\gamma}$ and inversely proportional to $h^2$.
The latter agrees with the finding in a quantum mechanical study
of small spin baths \cite{beuge16}, which supports the 
assumption that pulse model II corresponds better to the pulsing of the quantum mechanical 
model. The data collapse obtained by the scalings is not quantitative due
to the transition and the concomitant non-monotonic dependence on system size
and magnetic field.

\section{Results for the central spin model with nuclear Zeeman coupling}
\label{sec:WNZ}
 
In any experiment, an applied external magnetic field acts on the electronic spin
as well as on the nuclear spins by the Zeeman effect. In many circumstances, the latter
can safely be neglected because it is smaller by three orders of magnitude due
to the larger mass of the nuclei compared to the mass of electrons. But in the CSM
as a model for quantum dots, a magnetic field of $2\,\mathrm{T}$ is about two orders of magnitude
larger than the intrinsic energy scale $\jq$ of the CSM. For the large spin baths
with $\gamma=10^{-5}$, the electronic magnetic field is four orders of
magnitude larger than the largest individual coupling $J_k$ to a single nuclear spin.
Hence for each nuclear spin, the nuclear Zeeman effect is about ten times larger
than its coupling to the electronic (central) spin. Thus, this effect needs
to be considered \cite{beuge17} and it is possible that it introduces
qualitatively important changes in the dynamics; for instance, there is
evidence that even resonance is favored over odd resonance by including
the nuclear Zeeman effect \cite{beuge17,jasch17}.

We include the nuclear Zeeman coupling in the Hamiltonian \eqref{eq:hamil} by setting 
$z = 1/800$. First, we analyze periodic pulsing with pulse model I, and then we analyze periodic pulsing
with pulse model II.

\subsection{Pulse model I from Eq. \eqref{eq:pulse1}}
\label{subsec:WNZ_1}

\begin{figure}[htb]
	\centering
	\includegraphics[width=1.0\columnwidth]{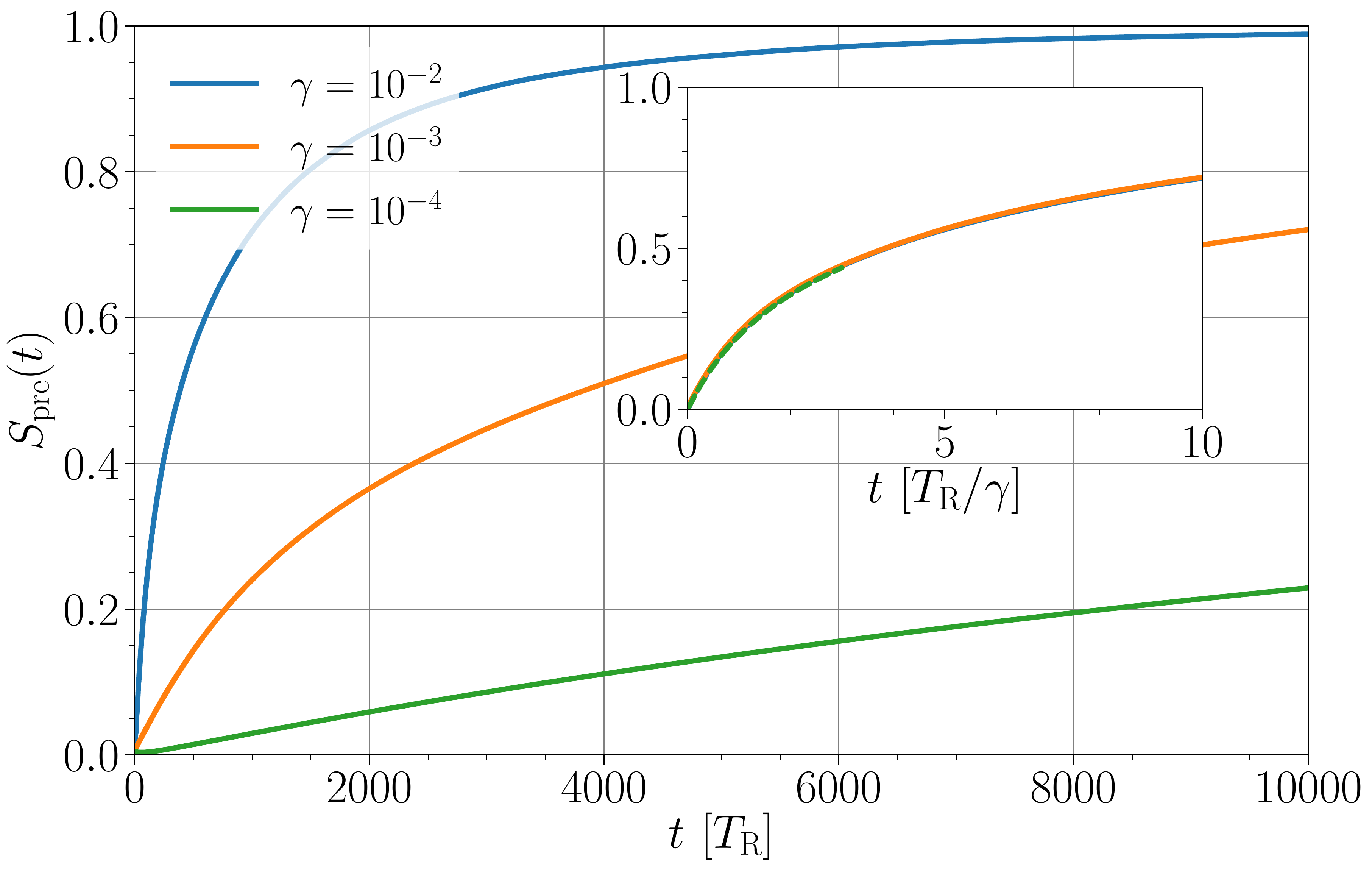}
	\caption{
		Relative pre-pulse signal as a function of time at $h=40\jq$ and pulse repetition time
		$T\R=5\pi/\jq$ for various sizes $2/\gamma$ of the spin bath 
		using pulse model I from \eqref{eq:pulse1} including the nuclear Zeeman effect. 
		Inset: Excellent data collapse when the time axis is scaled with $\gamma$.}
	\label{fig:prepulse_wnz_1}
\end{figure}

The relative pre-pulse amplitude $S_\mathrm{pre}(t)$ due to 
periodic pulsing with pulse model I is shown in Fig.\ \ref{fig:prepulse_wnz_1}.
The maximum pre-pulse signal $S_\mathrm{pre}(t \to \infty) = 1$ 
is approached for $\gamma = 10^{-2}$
while the curves for smaller values of $\gamma$ have not yet reached saturation.
A comparison to the curves without nuclear Zeeman splitting in Fig.\ \ref{fig:prepulse_1}
reveals that the nuclear Zeeman effect slows down the build-up of the pre-pulse signal
significantly, especially for smaller values of $\gamma$.

It is obvious that the scaling proportional to $\sqrt{\gamma}$
will not work anymore. Instead, we find a remarkable data collapse 
scaling the time with $\gamma$ as depicted in the inset of Fig.\ \ref{fig:prepulse_wnz_1}.
This finding is in agreement with what has been observed by J\"aschke et al.\
very recently in a semi-classical analysis. They found a data collapse 
by scaling the time with the inverse size of the spin bath \cite{jasch17}.

The question arises as to why the scaling changes from $\sqrt{\gamma}$ to $\gamma$
upon including the nuclear Zeeman effect. We attribute this qualitative change 
to the relative strengths of the couplings to which an individual bath spin
is subjected. Without nuclear Zeeman effect the coupling $J_k$ is the only
energy, hence rate, relevant for the individual bath spin. These couplings
scale like $\sqrt{\gamma}$ and thus, the evolution of the central spin
exerts an effect onto each bath spin at a rate $\propto \sqrt{\gamma}$.

But if the nuclear Zeeman effect with $zh$ is considered,
each bath spin is dominated by this term and precesses about the external
field. Then the coupling to the central spin is just
a perturbation on top of the coupling to the
external magnetic field. This perturbation is effective only in second order $J_k^2/zh$.
This difference is similar to the Stark effect which is generically second order,
but first order if the perturbed system is degenerate, i.\,e., without internal
dynamics. With nuclear Zeeman effect, the central spin dynamics
influences the bath spins only in second order.

In practice, we use the scaling and perform the following calculations
for relatively large values of $\gamma = 10^{-2}$ because no qualitative
changes for smaller values are to be expected. We stress that for realistic values 
$\gamma=10^{-5}$ all time dependences are slower by three orders of
magnitude.

\begin{figure}[htb]
	\centering
	\includegraphics[width=1.0\columnwidth]{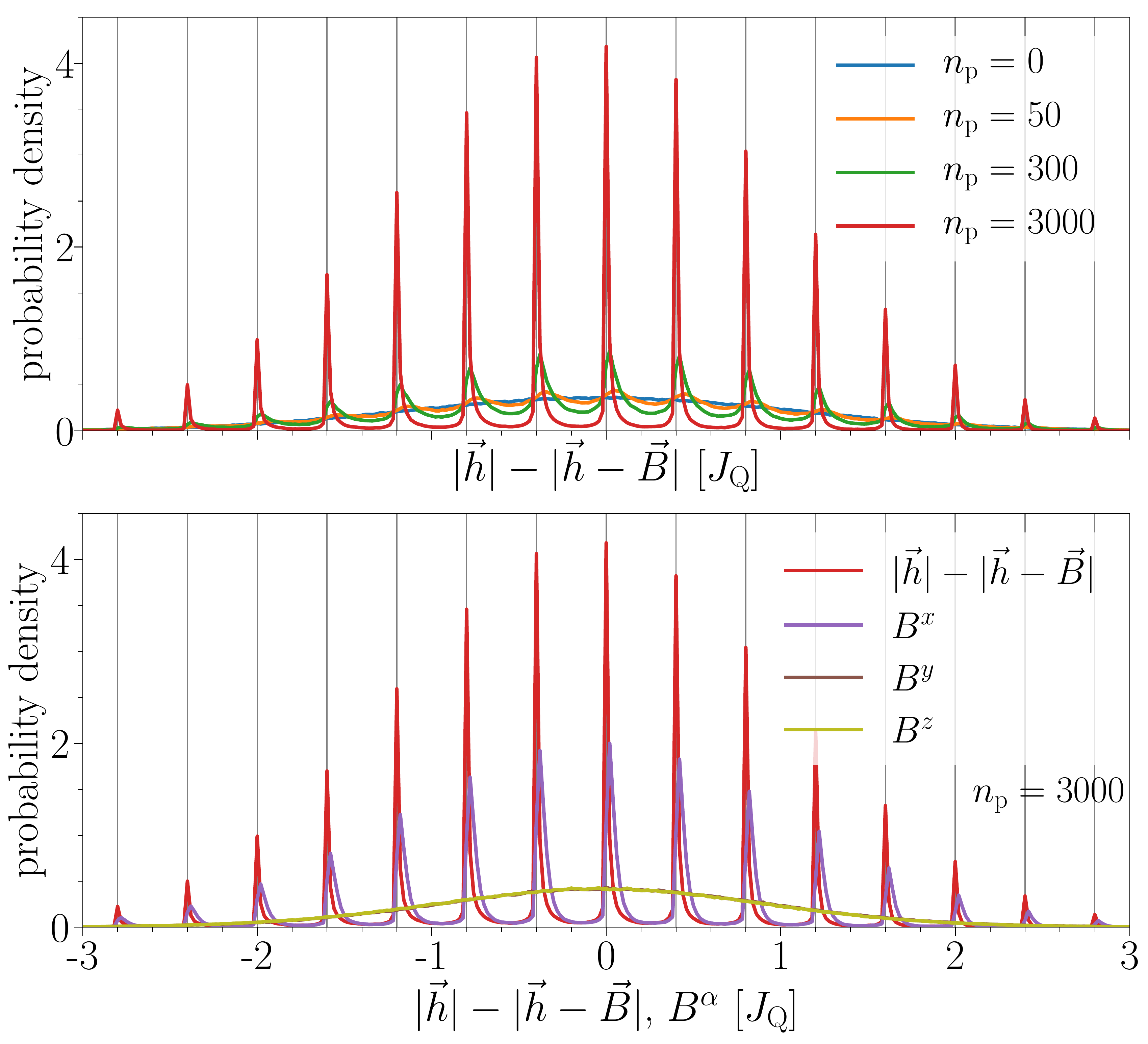}
	\caption{Upper panel: Distribution of the effective Larmor frequency $|\vec{h} - \vec{B}|$, shifted by $|\vec{h}|$, at $h=40\jq$ and pulse repetition time $T\R=5\pi/\jq$ 
	for $\gamma = 10^{-2}$ after $n_\mathrm{p}$ pulses using pulse model I from \eqref{eq:pulse1} and including the nuclear Zeeman effect. The vertical solid lines indicate the 
	values which satisfy the even resonance condition \eqref{eq:even}.
	Lower panel: Same as upper panel, but after $n_\mathrm{p} = 3000$ pulses. Additionally, 
	the distributions of the Overhauser field components $B^\alpha$ ($\alpha\in\{x,y,z\}$) are shown.}
	\label{fig:Overhauser_wnz_1}
\end{figure}

A representative distribution of effective Larmor frequency is plotted in the upper panel of Fig.\ 
\ref{fig:Overhauser_wnz_1}. The peaks are located at the values of $|\vec{h} - \vec{B}|$ 
fulfilling the even resonance condition \eqref{eq:even} in contrast to
what we found for the same pulse without nuclear Zeeman effect in
Sect.\ \ref{subsec:NNZ_1}. The fact that the nuclear Zeeman effect
favors the even resonance is in line with previous evidence \cite{beuge17,jasch17}.
Additionally, we observe a certain asymmetry with tails to large
values of the Overhauser field. For long pulsing the peaks become
sharper and sharper and nuclear focusing appears to become
perfect, i.\,e., the peaks in the distribution of the effective Larmor frequency
approach $\delta$-peaks. This is strongly corroborated by the evolution
of the relative pre-pulse signal which approaches unity (see Fig.\ \ref{fig:prepulse_wnz_1}).

The distributions of $B^y$ and $B^z$ maintain their initital Gaussian shape.
The inclusion of the nuclear Zeeman effect leads to a drastically different 
behavior as we find no finite polarization $\overline{B^y}>0$ anymore.
Again, the peaks of in the distribution of $B^x$ are slightly shifted to the 
right of the theoretical resonance conditions since the variances of $B^y$ and 
$B^z$ contribute to the effective Larmor frequency.

\begin{figure}[htb]
	\centering
	\includegraphics[width=1.0\columnwidth]{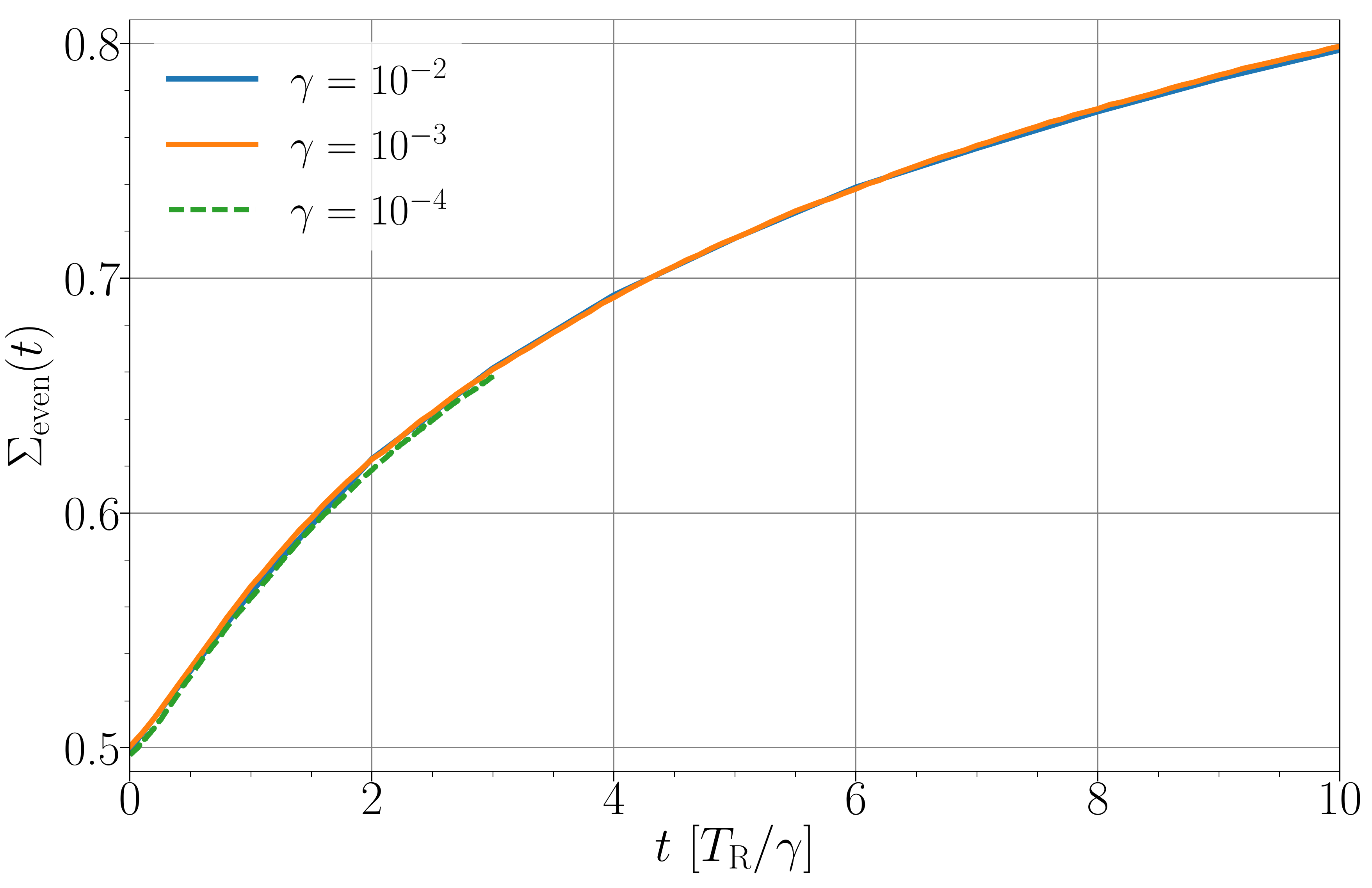}
	\caption{Weight $\Sigma_\mathrm{even}$ of the even resonances in the 
	Overhauser field distribution as a function of scaled time $t\gamma$ at $h=40\jq$ 
	and pulse repetition time $T\R=5\pi/\jq$ for various size of the spin bath
	using pulse model I from  \eqref{eq:pulse1} including the nuclear Zeeman effect.}
	\label{fig:scaled_weight_gamma_wnz_1}
\end{figure}

Further strong support for perfect nuclear focusing is provided by the study
of the weight of the even resonances shown in Fig.\ 
\ref{fig:scaled_weight_gamma_wnz_1} versus the time scaled by $\gamma$.
For long times, the even weight approaches unity, which means
that all Overhauser fields evolve towards values compatible with the
even resonance condition \eqref{eq:even}.
In addition, the scaled curves of Fig.\ 
\ref{fig:scaled_weight_gamma_wnz_1} yield a remarkably perfect data collapse
corroborating the scaling with $\gamma$.

\begin{figure}[htb]
	\centering
	\includegraphics[width=1.0\columnwidth]{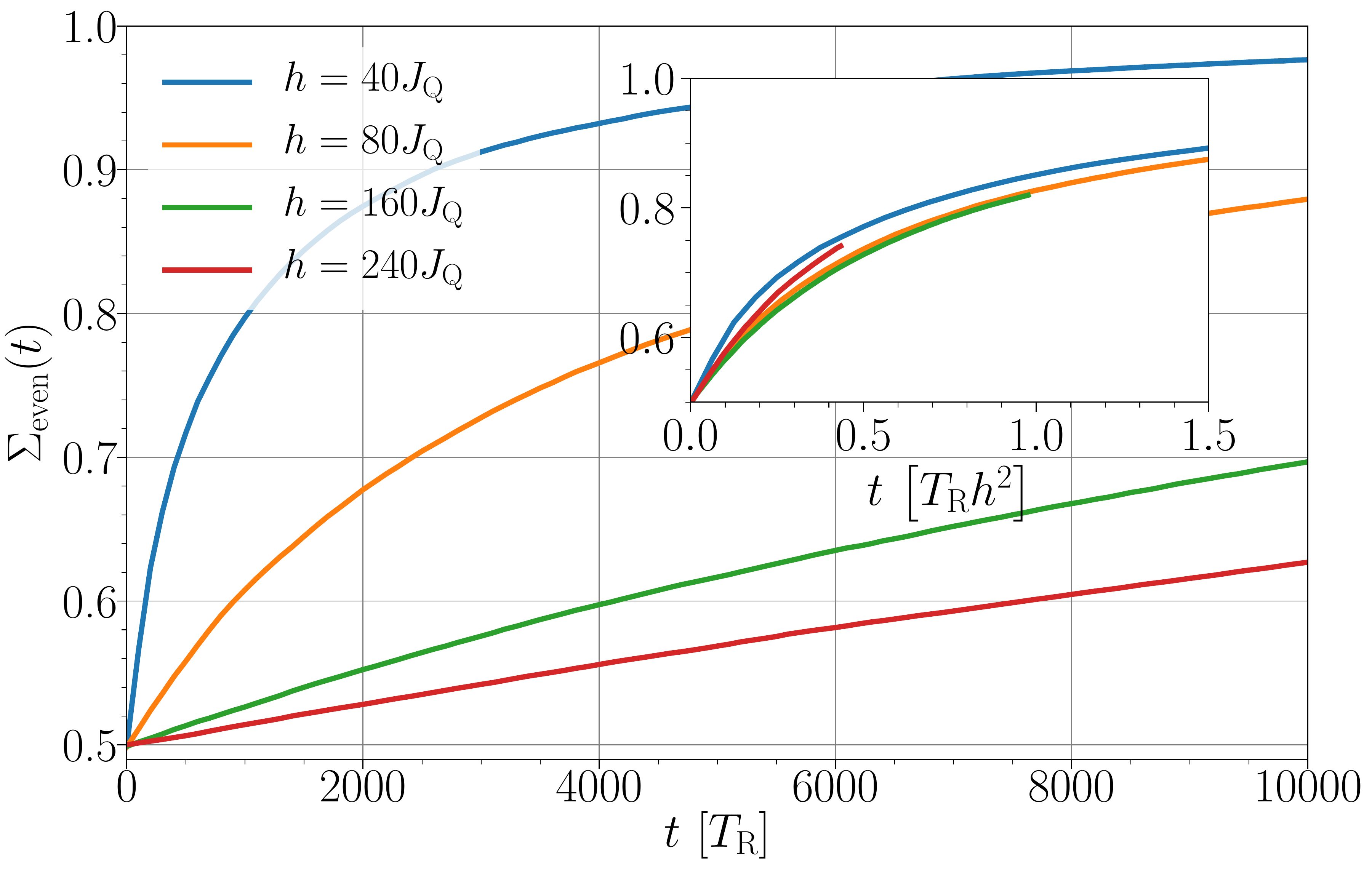}
	\caption{Weight $\Sigma_\mathrm{even}$ of the Overhauser field distribution fulfilling the even resonance
	condition \eqref{eq:even} as a function of time for $\gamma = 10^{-2}$ and pulse repetition 
	time $T\R=5\pi/\jq$ for various external magnetic fields $h$ using pulse model I 
	from \eqref{eq:pulse1} and including the nuclear Zeeman coupling. Inset: Time axis scaled with $1/h^2$.}
	\label{fig:weight_h_wnz_1}
\end{figure}

Next, we turn to varying the magnetic field. In Fig.\ \ref{fig:weight_h_wnz_1}
the increasing even weight is plotted for various magnetic fields. Clearly,
larger fields result in slower changes in the Overhauser field distribution.
The best collapse of the curves is obtained for scaling the time proportional
to $1/h^2$ as illustrated nicely in the inset of Fig.\ \ref{fig:weight_h_wnz_1}.
We emphasize that the scaling is quadratic with the magnetic field in contrast
to what we found in Sect.\ \ref{subsec:NNZ_1} for periodic pulsing with pulse model I
without nuclear Zeeman effect. We attribute this change of scaling
to the fact that the individual change of each bath spin has become
a second-order effect due to the precession of the nuclear spins
about the external magnetic field. Hence, the Overhauser field has become stiffer
due to the nuclear Zeeman effect.

Because the nuclear Zeeman coupling introduces an additional time scale which depends on the external field $h$, it is reasonable to expect a qualitative change of the physics when increasing 
$h$ from $40\jq$ to $240\jq$. A conceivable scenario would be a transition from even to odd resonance. However, we notice no qualitative difference for the studied values of $h$ as shown by Fig.\ \ref{fig:weight_h_wnz_1}.
Instead, the inset in Fig.\ \ref{fig:weight_h_wnz_1} indicates that the dynamics of the weight 
$\Sigma_\mathrm{even}(t)$ scales with $1/h^2$. 

For completeness, we also studied the phase jumps around the pulses.
They are found to be close to zero, $\Delta \varphi \approx 0$, as expected
for dominant even resonance. So we obtained a complete, consistent
picture of the CSM subject to pulses of type I including the nuclear Zeeman
effect. 

In summary, we find very strong nuclear focusing triggered by periodic 
application of pulse model I including the nuclear Zeeman effect. 
The inclusion of the nuclear Zeeman effect
slows the rates of nuclear focusing down considerably. The scaling with
the inverse spin bath size $\gamma$ changes from $\sqrt{\gamma}$ to $\gamma$. 
The scaling with magnetic field changes from $1/h$ to $1/h^2$. 
These scaling laws lead to very good data collapse, i.\,e., the scaling
is quantitative for pulse model I.

\subsection{Pulse model II from Eq. \eqref{eq:pulse2}}
\label{subsec:WNZ_2}

\begin{figure}[htb]
	\centering
	\includegraphics[width=1.0\columnwidth]{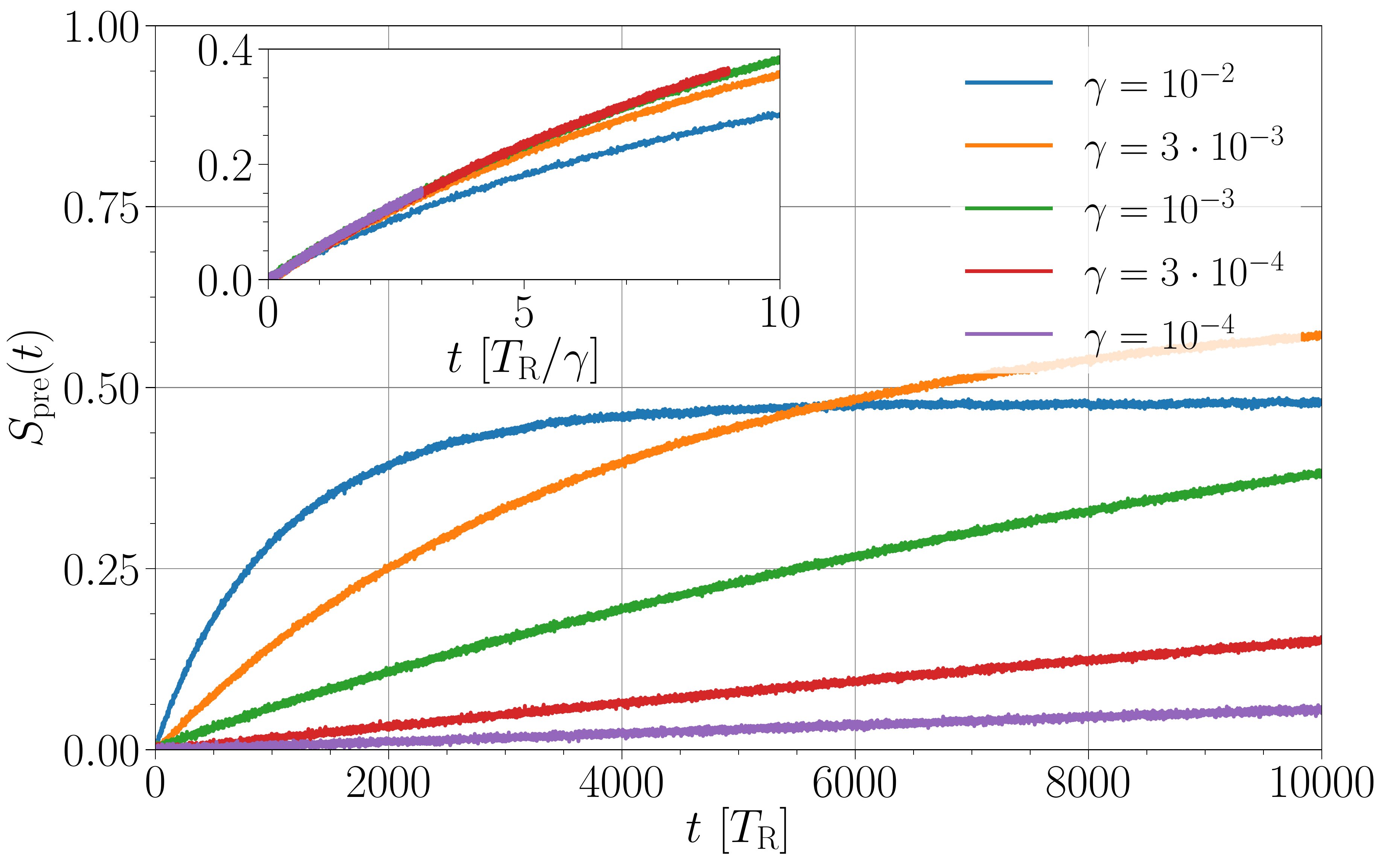}
	\caption{
		Relative prepulse signal as a function of time at $h=40\jq$ and pulse repetition time
		$T\R=5\pi/\jq$ for various spin baths using pulse model II from \eqref{eq:pulse2} and including the nuclear Zeeman coupling.
		Inset: Time axis scaled with $\gamma$.}
	\label{fig:prepulse_wnz_2}
\end{figure}

Next, we study the periodic pulsing by pulse model II in presence of the nuclear Zeeman
effect. Again, we start by inspecting the relative pre-pulse amplitude 
$S_\mathrm{pre}(t)$ for various values of $\gamma$ in Fig.\ \ref{fig:prepulse_wnz_2}.
Compared to the case without nuclear Zeeman coupling (see Fig.\ \ref{fig:prepulse_2}), the pre-pulse
signal is much more pronounced. But we stress that still perfect saturation is not reached,
i.\,e., the relative pre-pulse signal becomes stationary,
but its value remains significantly below its theoretical
maximum of unity: $S_\mathrm{pre}(t\to\infty) < 1$ (see blue curve in Fig.\ \ref{fig:prepulse_wnz_2}).
In this sense, nuclear frequency focusing remains imperfect.

\begin{figure}[htb]
	\centering
	\includegraphics[width=1.0\columnwidth]{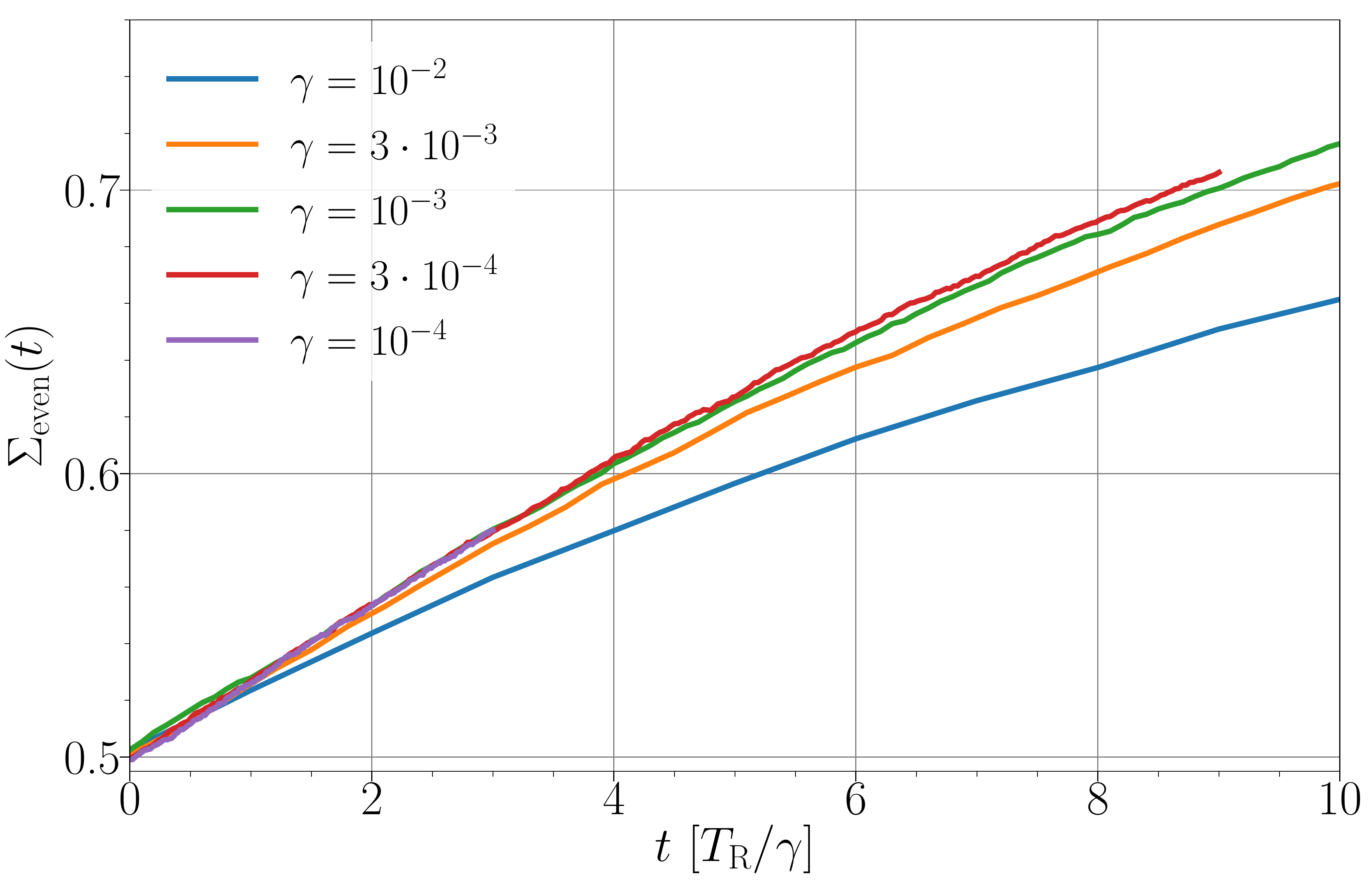}
	\caption{Weight $\Sigma_\mathrm{even}$ of the Overhauser field distribution fulfilling the even resonance condition as a function of the scaled time $t\gamma$ at $h=40\jq$ and pulse repetition time $T\R=5\pi/\jq$ for various spin baths using pulse model II from \eqref{eq:pulse2} and including the nuclear Zeeman coupling. }
	\label{fig:scaled_weight_gamma_wnz_2}
\end{figure}

A scaling of the time axis with $\gamma$ as depicted in the inset of Fig.\ \ref{fig:prepulse_wnz_2} 
appears to work for very small values of $\gamma \ll 10^{-2}$ for the relative pre-pulse
signal. But the data collapse is not as good as 
for pulse \eqref{eq:pulse1} (see Fig.\ \ref{fig:prepulse_wnz_1}).
The weight $\Sigma_\mathrm{even}(t)$ shows a very similar behavior as
to be expected for consistency (see Fig.\ \ref{fig:scaled_weight_gamma_wnz_2}).

\begin{figure}[htb]
	\centering
	\includegraphics[width=1.0\columnwidth]{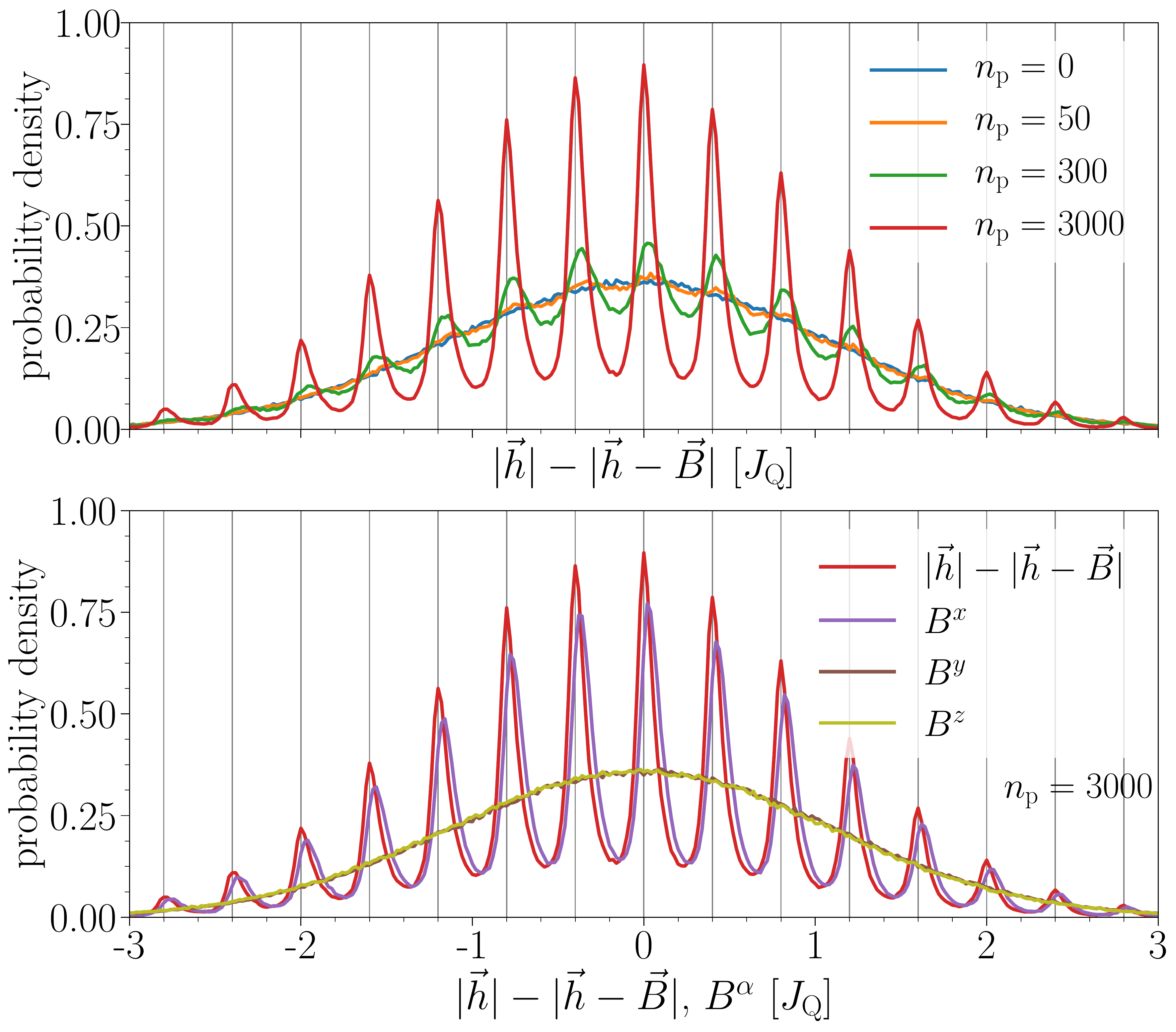}
	\caption{Upper panel: Distribution of the effective Larmor frequency $|\vec{h} - \vec{B}|$, shifted by $|\vec{h}|$, at $h=40\jq$ and pulse repetition time $T\R=5\pi/\jq$ 
	for $\gamma = 10^{-2}$ after $n_\mathrm{p}$ pulses using pulse model II from \eqref{eq:pulse2} and including the nuclear Zeeman effect. The vertical solid lines indicate the 
	values which satisfy the even resonance condition \eqref{eq:even}.
	Lower panel: Same as upper panel, but after $n_\mathrm{p} = 3000$ pulses. Additionally, 
	the distributions of the Overhauser field components $B^\alpha$ ($\alpha\in\{x,y,z\}$) are shown.}
\label{fig:Overhauser_wnz_2}
\end{figure}

Representative results for the distributions of the effective Larmor frequency distribution and for the Overhauser field components are presented in Fig.\ 
\ref{fig:Overhauser_wnz_2}. The Overhauser field components $B^y$ and $B^z$ maintain their initial Gaussian shape. 
The peaks in the distribution of $B^x$ are again slightly shifted to the right.
The peaks and therefore the degree of nuclear focusing are noticeably more strongly pronounced than for the case without nuclear Zeeman coupling (Fig.\ \ref{fig:Overhauser_2}).
The peak positions of the effective Larmor frequency fulfill the even resonance condition \eqref{eq:even}, which is also visible on inspecting the weight $\Sigma_\mathrm{even}(t)$ in Fig.\ 
\ref{fig:scaled_weight_gamma_wnz_2}.
The scaling with $\gamma$ works as well if $\gamma$ is chosen small enough.

\begin{figure}[htb]
	\centering
	\includegraphics[width=1.0\columnwidth]{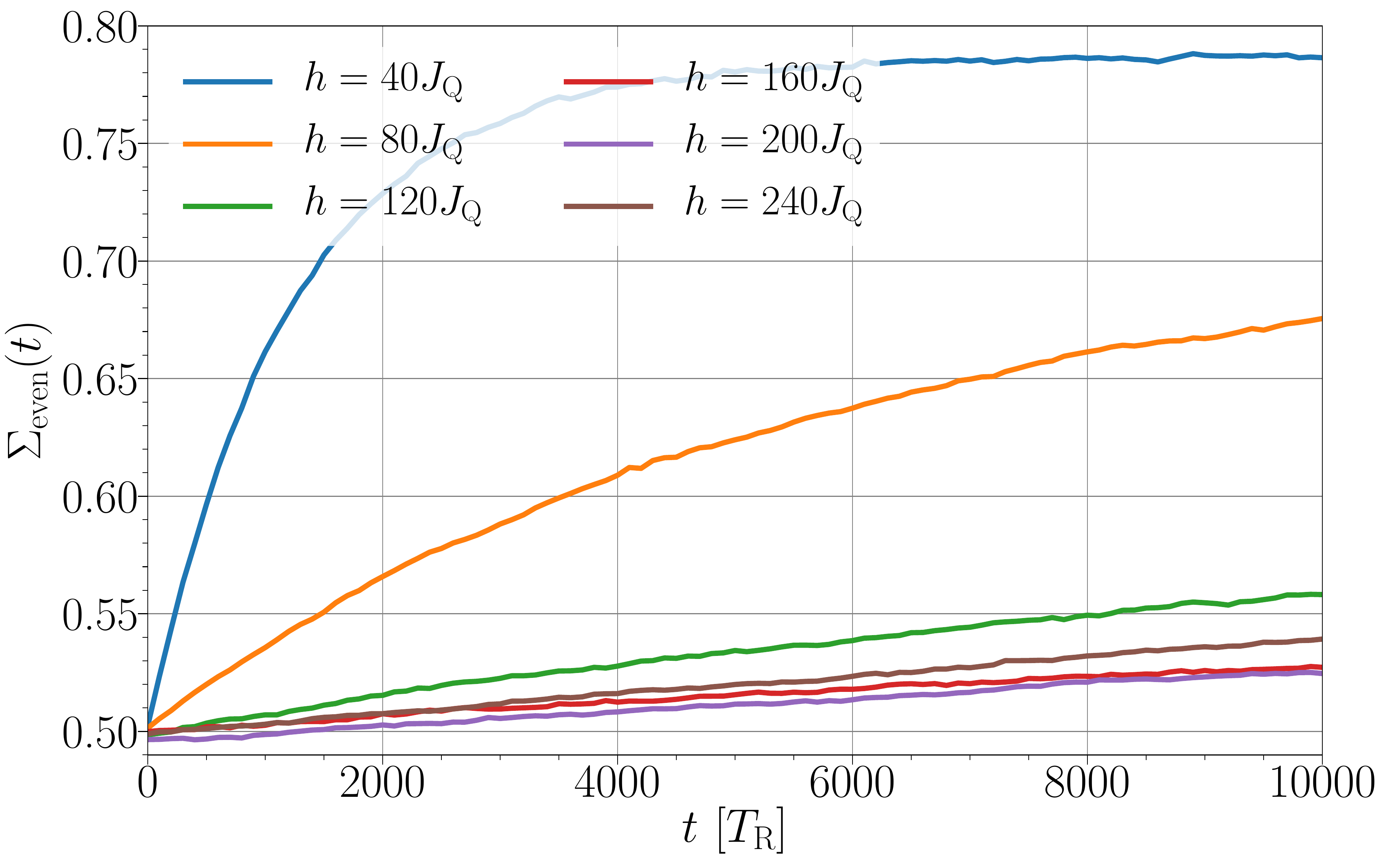}
	\caption{Weight $\Sigma_\mathrm{even}$ of the Overhauser field distribution fulfilling the even resonance condition as a function of time for $\gamma = 10^{-2}$ at pulse repetition time 
	$T\R=5\pi/\jq$ for various external magnetic fields $h$ using pulse model II from \eqref{eq:pulse2} and including the nuclear Zeeman coupling.}
	\label{fig:weight_h_wnz_2}
\end{figure}

The dependence of $\Sigma_\mathrm{even}$ on the magnetic field $h$ 
for fixed $\gamma = 10^{-2}$ and $10^{-3}$ is shown in Fig.\
 \ref{fig:weight_h_wnz_2}.
Again, larger values of $h$ imply a slower build-up of nuclear focusing.
In contrast to the previous results, we find no perfect data collapse by either scaling
with $1/h$ or with $1/h^2$. Additionally, we only find even resonance in all calculations
done in this subsection independent of the choice of $\gamma$ and $h$. 
This conclusion is supported by the weight $\Sigma_\mathrm{even} > 0.5$
and by the vanishing phase differences $\Delta \varphi \approx 0$.

\begin{figure}[htb]
	\centering
	\includegraphics[width=1.0\columnwidth]{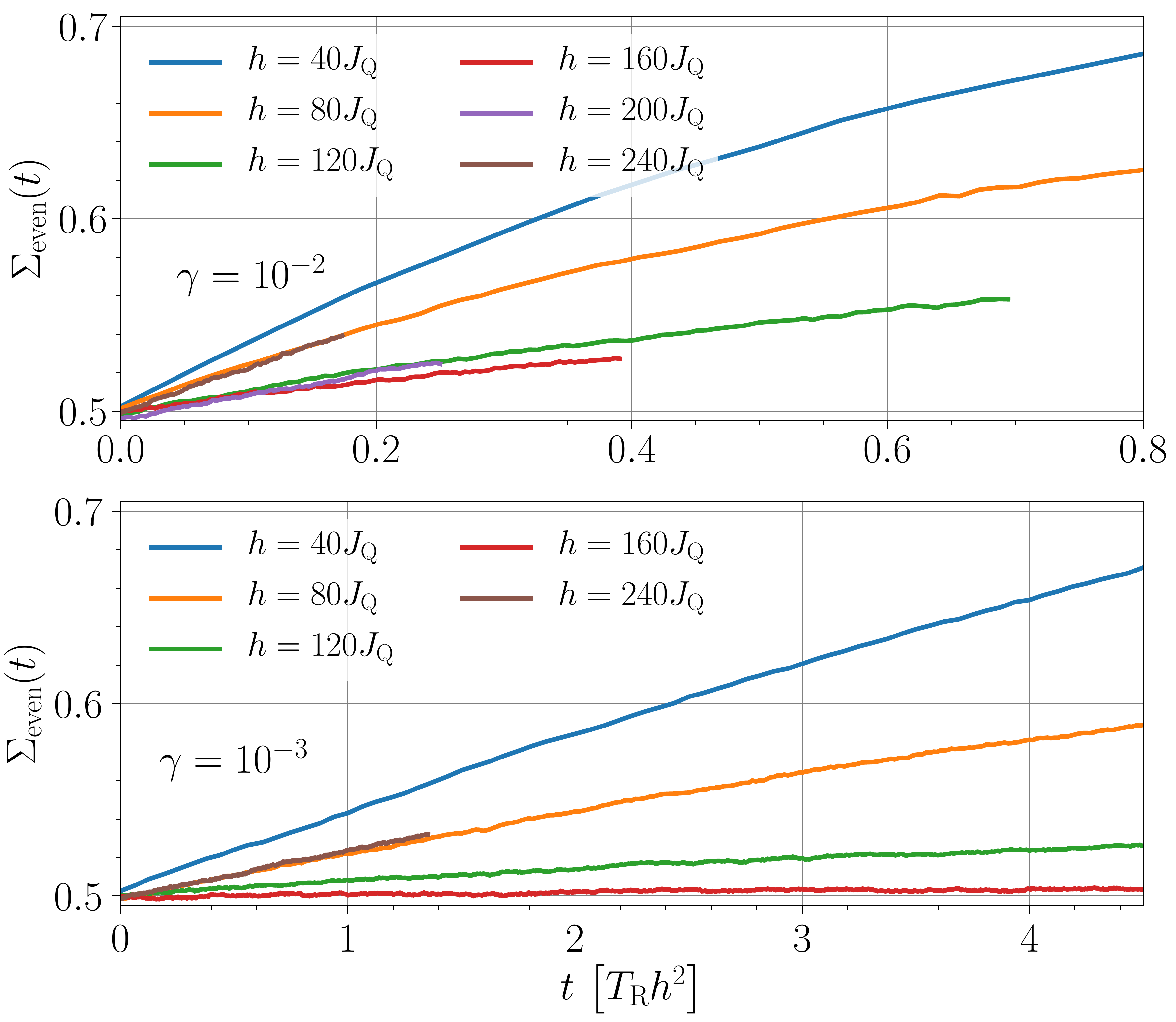}
	\caption{Weight $\Sigma_\mathrm{even}$ of the Overhauser field distribution fulfilling the even resonance condition as a function of the scaled time $t/h^2$
	for $\gamma = 10^{-2}$ and pulse repetition time $T\R=5\pi/\jq$ for various external magnetic fields $h$ using pulse model II from \eqref{eq:pulse2} and including the nuclear Zeeman coupling. Taking the scale on the time axis into account one notes 
	the significantly slower dynamics for lower values of $\gamma$ in the lower panel.}
	\label{fig:scaled_weight_h_wnz_2}
\end{figure}

Interestingly, the rate of the build-up of  nuclear focusing 
shows a minimum at around $h=160\jq$, see Figs.\ \ref{fig:weight_h_wnz_2} 
and \ref{fig:scaled_weight_h_wnz_2}. This is clearly discernible if the
time is scaled with $1/h^2$ as in Fig.\ \ref{fig:scaled_weight_h_wnz_2}. 
This scaling does not yield a perfect data collapse, but
it is obvious that it captures the main effect of the magnetic field
on the time evolution rates, except for the non-monotonic behavior.

The minimum appears to be more pronounced if the effective bath size is increased, 
i.\,e., if $\gamma$ is decreased.
In physical units, the minimum at around $h=160\jq$ corresponds to a field of roughly $4 \,\mathrm{T}$. There may be a connection to recent experiments because a minimum of the 
pre-pulse signal has been found at $3.75\,\mathrm{T}$ in Ref.\ \onlinecite{jasch17}. 
But the present analysis is not yet clear enough to draw definite conclusions.
The experimental data appear to be in the saturated stationary state at 
very long times while the numerical data are not.
Unfortunately, we cannot reach saturation in our simulation yet 
for the large values of $h$ due to the lack of computational resources. 

In summary, we find  significant nuclear focusing for pulse model II, but less
pronounced than for pulse model I. The inclusion of the nuclear Zeeman effect
slows down the rates of nuclear focusing considerably. The scaling with
the inverse spin bath size 
$\gamma$ changes from $\sqrt{\gamma}$ to $\gamma$. The scaling
with magnetic field remains $1/h^2$, i.\,e., there is no
change upon inclusion of the nuclear Zeeman effect. Remarkably, the nuclear focusing
induced by pulse model II displays non-trivial non-monotonic features, but the even resonance
clearly prevails. The non-monotonicity prevents a perfect data collapse upon scaling.

\section{Conclusions}
\label{sec:conclusion}

Many experiments showed that periodically pulsed electronic spins in ensembles of
quantum dots display nuclear focusing such that the Larmor precessions synchronize
to the periodicity of the external pulses. Here we simulated this physical setup
by classical pulses applied to a classical central spin model. Recent algorithmic
progress makes it possible to simulate very large spin baths which reach
the experimentally relevant sizes of up to $10^5$ bath spins.

We studied two kinds of pulses which both align the central spin along the $z$-direction. 
Pulse model I aligns the total spin vector while pulse model II keeps knowledge of the 
quantum mechanical uncertainty so that the transversal components remain finite.
Both kinds of pulses have been applied to an isotropic central spin model without and with the
nuclear Zeeman term.

In all cases, we found strong signatures of nuclear frequency focusing. This is signalled by
a strong pre-pulse signal of the central electron spin, i.\,e., a signal similar to a spin echo which occurs 
\emph{before} the next pulse is applied. Perfect nuclear focusing leads to a 
saturated pre-pulse signal which is as large as the signal right after the pulse.
This phenomenon is explained by a highly non-equilibrium distribution of the
Overhauser field, i.\,e., the effective magnetic field exerted by the ensemble
of nuclear bath spins. Its distribution develops a comb-like peak structure such
that the difference between the Overhauser fields in two adjacent peaks implies 
precisely one additional {spin revolution} between two pulses 
\cite{greil06a,greil06b,greil07a,petro12,glazov12,beuge16,beuge17,jasch17}.
We distinguish between odd and even resonances in the peaks of the Overhauser distribution.
In an even resonance, an integer number of {revolutions} takes place between two
consecutive pulses. In an odd resonance, a half-integer number of 
{revolutions} takes place.

The nuclear focusing induced by pulse model I is very efficient. Odd resonances occur
and the build-up of the pre-pulse signal scales with $1/\sqrt{\neff}$, i.\,e.,
if the spin bath is four times larger the nuclear focusing takes place slower
by a factor of 2 pre-supposing that the short time dynamics is the same.
Similarly, larger magnetic fields slow down nuclear focusing in a linear fashion in $1/h$.
These scalings yield a very good data collapse so that quantitative extrapolations
are possible.

For pulse model II without the nuclear Zeeman term, 
nuclear focusing can arise in odd or in even resonance with a transition between
the two scenarios. The typical experimental numbers are such that the odd resonance
is the relevant one. The degree of nuclear focusing is always weaker than for pulse
I, i.\,e., the pre-pulse signal does not reach perfect saturation and the peak structure
in the distribution of the Overhauser field does not approach a comb of 
$\delta$-functions, but the peaks retain a certain width.
The scaling of the build-up of nuclear focusing is again $\propto 1/\sqrt{\neff}$,
but it scales with $1/h^2$ in magnetic field in accordance with the quantum
mechanical finding \cite{beuge16}. The scalings yield an approximate data collapse only
due to the non-monotonic dependence on the parameters and the transition between
odd and even resonance.

Next, we included the nuclear Zeeman term with realistic average values.
For pulse model I, we observed that the nuclear focusing shifts from odd resonance 
to even resonance. Furthermore, the build-up is slowed down considerably due
to the nuclear Zeeman term. The scaling of the build-up rate is now proportional
to $1/\neff$ and to $1/h^2$. Still, the scaling yields a very good data collapse
enabling quantitative extrapolations.

Inclusion of the nuclear Zeeman term for periodic pulsing with pulse model II yields 
even resonances. Roughly, the rate of build-up of nuclear focusing is proportional to 
$1/\neff$. 
But due to the non-monotonic behavior {on the magnetic field $h$, no perfect data collapse can be obtained by scaling with $1/h^2$.  Still, the relevant rate of change
is proportional to $1/h^2$}. 

Interestingly, nuclear focusing {is 
only very weak around} a magnetic field of $4$T. A similar phenomenon 
has been observed in experiments \cite{jasch17}, but {further computationally}
 demanding calculations are required to establish the relation to experiment 
quantitatively.

Generally, we emphasize that our findings clearly show that a realistic
description of the pulse process matters. The differences between long trains
of pulse model I or pulse model II underline that a quantitative understanding requires us 
to know and to describe what the laser pulses do. Though the phenomenon
of nuclear focusing as such appears to be robust \cite{petro12}, important
features such as the speed of the build-up, the value of possible 
saturated pre-pulse signals, and the nature of the resonance (even or odd)
do depend on the nature of the pulses.

Finally, we have pointed out that the Overhauser field can acquire also non-zero
transversal components. So far, it always turned out that the transversal
components remain Gaussian distributed with zero average values. But at least
for pulse model I and zero nuclear Zeeman effect we found that the full resonance
condition has to be taken into account. Otherwise, it may even happen
that even and odd resonance appear to be interchanged.
The effective Larmor frequency depends
on the total effective magnetic field built from the external magnetic field
and the full Overhauser field. It would be very interesting to check experimentally
whether such transversal magnetizations play a role.

As an outlook, we stress that the model can be amended in several respects.
A first straightforward extension is to treat ensembles of quantum dots
with slightly varying $J_\mathrm{Q}$ and electronic $g$-factor which imply
inhomogeneous dephasing of the total signal stemming from all quantum dots
which is closer to many experiments.

Second, one can treat the pulse in a more realistic fashion by dealing with the
density matrix of the electronic spin and excited trion states. The 
spin bath exerts a classical Overhauser field, but the central spin 
is replaced by the expectation value of the spin operators. This mean-field
treatment captures a number of quantum aspects of the central spin and 
can be seen as a next step towards a realistic modeling of 
periodically driven spins in quantum dots.

Third, one can change the isotropic central spin model to an anisotropic one
which describes the physics of doped holes \cite{teste09,hackm14a}. In addition,
the different nuclear spins and their differing $g$-factors can be built-in
as well \cite{coish09,beuge17}. Thus, the present comprehensive paper paves the
way to many further steps towards a quantitative understanding of 
spins in semiconductor nanostructures.

\begin{acknowledgments}
We thank Frithjof B.\ Anders, Manfred Bayer, Vasilii Belykh, Wouter Beugeling, Eiko Evers, 
Mikhail Glazov, Lars B.\ Gravert, Alexander Greilich, Natalie J\"aschke, Robin R\"ohrig,
and Dmitry Smirnov for many helpful discussions. 
This paper has been supported financially by the Deutsche Forschungsgemeinschaft and 
the Russian Foundation for Basic Research in International Collaborative Research Centre TRR 160.
\end{acknowledgments}


\begin{appendix}

\section{Analysis of the phase jump to determine the resonance}
\label{App:phase}

\begin{figure}[htb!]
	\centering
	\includegraphics[width=1.0\columnwidth]{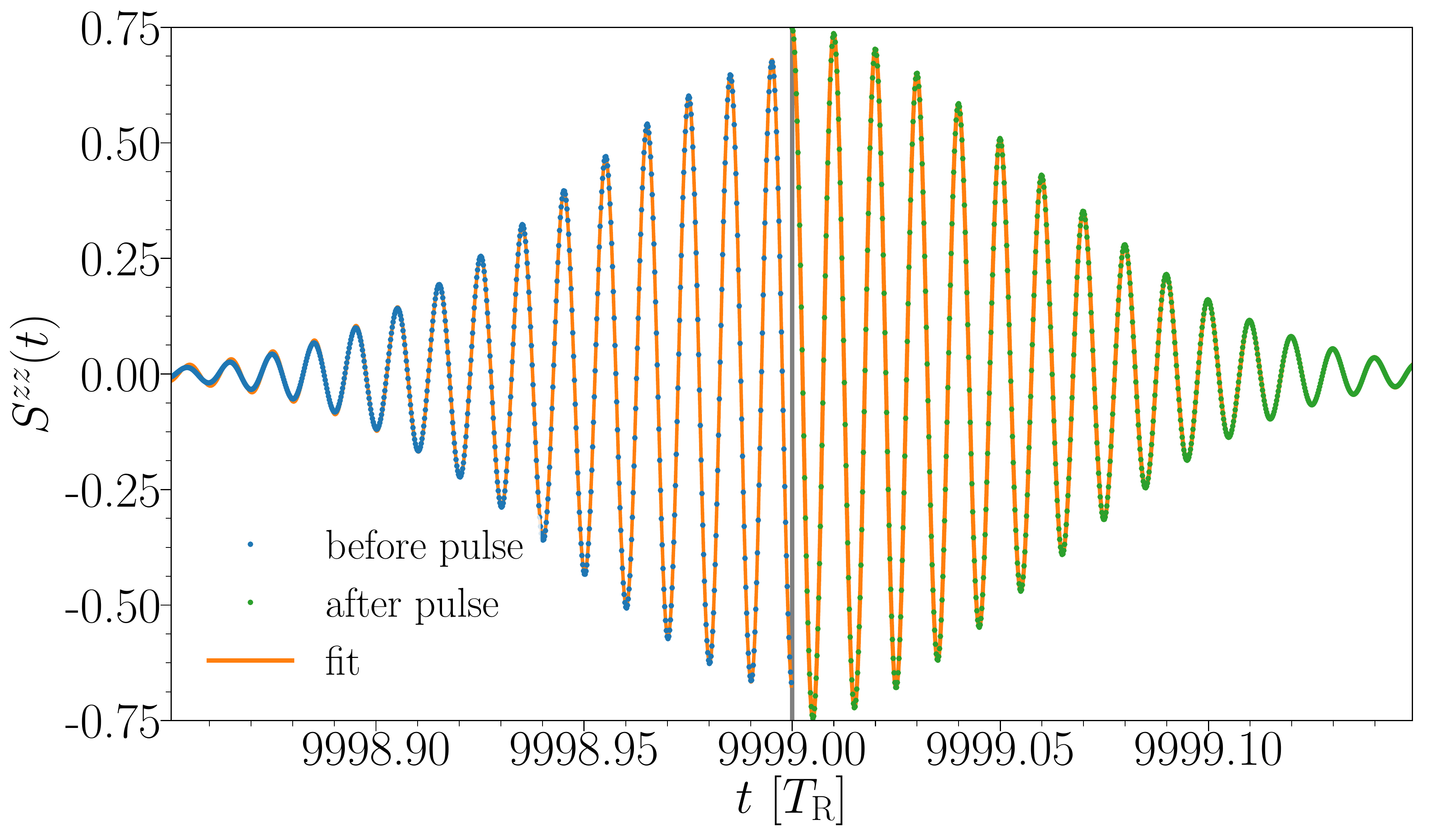}
	\caption{Example of the fit to obtain the phase jump $\Delta \varphi$; see
	main text and Eqs.\ \eqref{eq:fitfunction} and \eqref{eq:phasejump}.}
	\label{fig:phi_fit_1}
\end{figure}

We can calculate the phase jump $\Delta \varphi$ of the central spin precession by fitting a function of type
\begin{align}
f(t) = |A| \exp\left(- \frac{5 \jq^2 }{8} (t-t_0)^2 \right) 
\cos\left(ht - \varphi\right)
\label{eq:fitfunction}
\end{align}
to the autocorrelation function $S^{zz}(t)$ before and after the pulse separately.
The set of fit parameters is $A$, $t_0$, and $\varphi$. In general, they
will be different before and after the pulse.
The function $f(t)$ is chosen to comprise a Gaussian envelope modulating the amplitude of the 
Larmor precession with frequency $h$. Figure \ref{fig:phi_fit_1}
illustrates this kind of fit, displaying very nice agreement between the fit and the numerical data. 
Then, the phase jump $\Delta \varphi$ is defined by
\begin{align}
\label{eq:phasejump}
	\Delta \varphi = |\varphi_\mathrm{before} - \varphi_\mathrm{after}| \mod 2 \pi.
\end{align}

For pulse model I without nuclear Zeeman effect, we always find phase jumps of $\pi$ within
2 to 3\%.
This finding was checked for various $\gamma$ and external magnetic fields $h$.

For pulse model II without nuclear Zeeman effect, the phase jump takes values of zero or $\pi$ depending
on the combination of $\gamma$ and $h$. The corresponding resonance always matches the resonance found by analyzing the weight $\Sigma_\mathrm{even}$.
Table \ref{tab:phi_fit_2} provides some values for different combinations of 
$\gamma$ and $h$.
Note that for combinations of $\gamma$ and $h$ that are very close to 
$P = h \gamma^2 = 9 \cdot 10^{-5} \jq$, e.\,g., $\{h = 80 \jq,\, \gamma=1/\sqrt{2} 
\cdot 10^{-3}\}$, it is very hard to obtain a reliable value for $\Delta \varphi$
because the statistical error of the pre-pulse signal is of the order of the 
pre-pulse signal itself.

When including the nuclear Zeeman effect ($z = 1/800$), we always find $\Delta \varphi \approx 0$ for both pulse models I and II.

\begin{table}[htb!]
\caption{Phase jump $\Delta \varphi$ in units of $\pi$ 
		for various combinations of $\gamma$ and $h$ after $n_\mathrm{p}=10000$ pulses of type II without nuclear Zeeman effect.}
\begin{tabular}{|c||c|c|c|c|c|c|}
\hline
$h/\jq$ & 40 & 40 & 40  & 40 & 40 & 80 \\
\hline
$\gamma$ &  $10^{-2}$  & $3\cdot 10^{-3}$  
& $10^{-3}$ & $3\cdot 10^{-4}$ & $10^{-4}$  &  $3/\sqrt{2}\cdot 10^{-3}$ 
\\
\hline 
$\Delta\varphi/\pi$& 0.005  & 0.077   & 0.832  & 1.024 & 1.028 & 0.009
\\
\hline
\end{tabular}
\hspace{\fill}

\bigskip

\begin{tabular}{|c||c|c|c|c|c|}
\hline
$h/\jq$ & 80 & 80 & 160 & 240 \\
\hline
$\gamma$ &  $1/\sqrt{2}\cdot 10^{-3}$ & $1/(3\sqrt{2})\cdot 10^{-3}$  & $10^{-2}$ & $10^{-2}$
\\
\hline 
$\Delta\varphi/\pi$  &  1.237 & 1.027 & 0.005 & 0.011
\\
\hline
\end{tabular}
\hspace{\fill}
\label{tab:phi_fit_2}
\end{table}
\end{appendix}

\end{document}